\documentclass[aps,pre,longbibliography,notitlepage,floatfix,nofootinbib]{revtex4-1}

\usepackage{bm,amsmath,empheq,amssymb,graphicx,stmaryrd,float,subcaption,upgreek,sidecap,dsfont,mathtools,gensymb,multirow,array}
\usepackage[multiple]{footmisc}
\usepackage[abs]{overpic}
\usepackage{hyperref}

\usepackage{comment,footnote,mathrsfs,subcaption}
\usepackage[format=plain,labelfont={bf,small},textfont=small,justification=raggedright,singlelinecheck=false]{caption}

\usepackage{xcolor}

\newcommand{\bx}{{\bm x}}

\newcommand{\jump}[1]{{\left\llbracket #1 \right\rrbracket}} 
\newcommand{\avg}[1]{{\langle #1 \rangle}}

\newcommand{\bL}{{\bf L}}

\newcommand{\bO}{{\bm 0}}

\newcommand{\bB}{{\bm B}}
\newcommand{\bu}{{\bm u}}
\newcommand{\bl}{{\bm l}}
\newcommand{\bmo}{{\bm m}}

\newcommand{\bn}{{\bm n}}
\newcommand{\bP}{{\bf P}}

\newcommand{\bz}{{\bm z}}
\newcommand{\bF}{{\bf F}}

\newcommand{\bp}{{\bm p}}

\newcommand{\bd}{{\bm d}}
\newcommand{\be}{{\bm e}}

\newcommand{\tdelta}{{\tilde{\delta}}}

\usepackage[utf8]{inputenc}
\usepackage[T1]{fontenc}
\usepackage{nomencl}

\cleardoublepage           
\renewcommand{\nomname}{} 
\setcitestyle{numbers,square}

\begin{document}

\title{Planar equilibria of an elastic rod wrapped around a circular capstan}
\author{Harmeet Singh}
\email{harmeet.singh@epfl.ch, harmeet@vt.edu}
\affiliation{Laboratory for Computation and Visualization in Mathematics and Mechanics, Institute of Mathematics, \'Ecole Polytechnique F\'ed\'erale de Lausanne, 1015 Lausanne, Switzerland}
\date{\today}

\begin{abstract}
We present a study on planar equilibria of a terminally loaded elastic rod wrapped around a rigid circular capstan.
Both frictionless and frictional contact between the rod and the capstan are considered.
We identify three cases of frictionless contact -- namely where the rod touches the capstan at one point, along a continuous arc, and at two points.
We show that, in contrast to a fully flexible filament, an elastic rod of \emph{finite length} wrapped around a capstan does not require friction to support unequal loads at its two ends.
Furthermore, we classify rod equilibria corresponding to the three aforementioned cases in a limit where the length of the rod is much larger than the radius of the capstan.
In the same limit, we incorporate frictional interaction between the rod and the capstan, and compute limiting equilibria of the rod.
Our solution to the frictional case fully generalizes the \emph{classic capstan problem} to include the effects of finite thickness and bending elasticity of a flexible filament wrapped around a circular capstan.
\end{abstract}
\maketitle

\section{Introduction}
\begin{figure}[h!]
	\centering
		\includegraphics[width=0.95\linewidth]{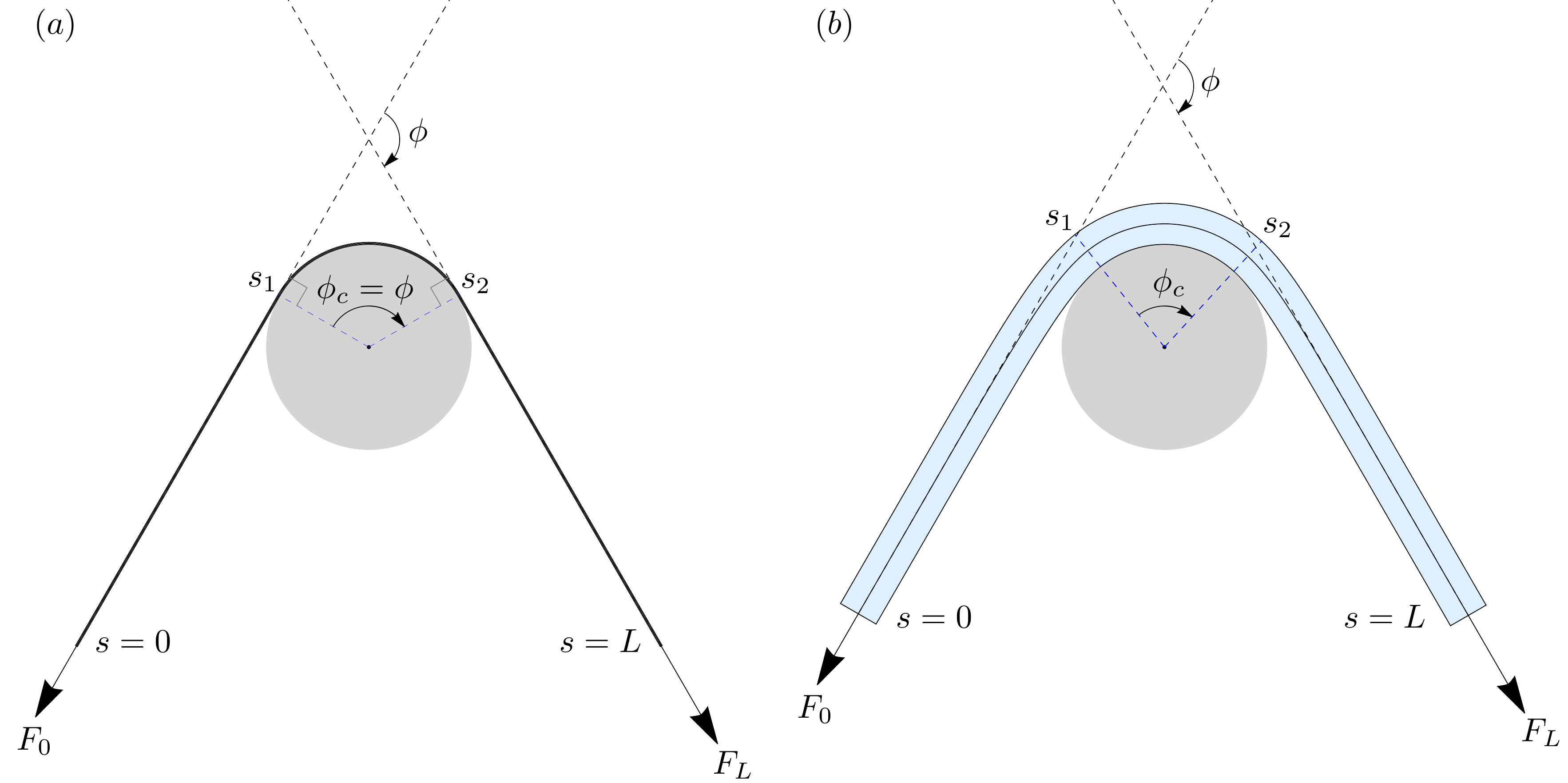}
	\caption{(a) A thin flexible filament with no bending elasticity wrapped around a circular capstan. The tangents at $s=0$ and $s=L$ remain aligned with the tangents at $s=s_1$ and $s=s_2$, respectively. This ensures that the contact angle $\phi_c$ equals the prescribed wrap angle $\phi$. (b) An elastic rod of finite thickness wrapped around a circular capstan. The bending elasticity of the rod causes the tangents at $s=s_1$ and $s=s_2$ to deviate from the tangents at $s=0$ and $s=L$, respectively. As a result, the contact angle $\phi_c$ has a complicated dependence on the wrap angle $\phi$, and the end loads $F_0$ and $F_L$.}
	\label{fig:introduction_schematics}
\end{figure}

The \emph{classic capstan problem} in mechanics comprises a fully flexible filament wrapped around a rigid circular capstan (Fig. \ref{fig:introduction_schematics}a), with the frictional interaction between them governed by Coulomb's inequality of static friction.
The problem entails computing the maximum possible output load $F_L$ that the filament can sustain at one end, for a prescribed input load $F_0$ at the other.
The two loads are related by the well known \emph{capstan equation}
\begin{align}
	F_L = F_0 e^{\mu\phi}\, ,\label{eq:classic_capstan}
\end{align}
where $\mu$ is the coefficient of static friction between the filament and the capstan, and $\phi$ is the wrap angle (Fig. \ref{fig:introduction_schematics}a) of the end loads.

The classic capstan problem is often introduced in undergraduate mechanics to demonstrate the role of friction in limiting equilibrium of flexible bodies \cite{maurer1944, meriam1978,hazelton1976,levin1991}.
In addition to its pedagogical significance, the problem has found relevance in several diverse domains of engineering, such as textile engineering \cite{howell1959,grosbergplate1968, mcgee1977,weichen1998,gaowanghao2015}, theory of power transmission by belts
 \cite{beflosky1973,firbank1970,beflosky1976,childs1980,kimmarshek1987,srivastavahaque2009,lubarda2015}, design of surgical robots \cite{baserkonukseven2010, xuerenyandu2017,simanwasinyang2018}, analysis of musical instruments \cite{groveskemp2019}, and even the mechanics of the DNA molecule \cite{ghosal2012}. 
Despite its wide use in engineering design with conservative values of the friction coefficient \cite{beflosky1973}, it has long been known that equation \eqref{eq:classic_capstan} is not adequately ratified by experiments \cite{howell1959,grosbergplate1968,mcgee1977}.
Workers in the field have typically attributed this discrepancy to primarily two reasons: i) most materials do not obey Coulomb's inequality of static friction \cite{lodgehowell1954}, and ii) the bending elasticity of the filament is completely ignored in the derivation of equation \eqref{eq:classic_capstan} \cite{grosbergplate1968}.
The main objective of this article is to remedy the latter shortcoming of the classical theory, by incorporating bending elasticity of the filament into the analysis.
 
Some of the earliest works on including bending elasticity of the filament in the capstan problem was done by I.M. Stuart \cite{stuart1961}. He recognized that bending elasticity would cause the filament to bow out near the contact boundaries (points $s_1$ and $s_2$ in Fig. \ref{fig:introduction_schematics}b).
Consequently, computing the contact angle $\phi_c$ for such a filament becomes a non-trivial task, unlike in the classic problem where $\phi_c$ is easily shown to be equal to the prescribed wrap angle $\phi$ of the end forces (Fig. \ref{fig:introduction_schematics}a).
Notwithstanding this realization, Stuart does not attempt to compute the unknown contact region (quantified by $\phi_c$) in \cite{stuart1961}, and instead formulates his theory with $\phi_c$ prescribed.
Subsequent attempts to generalize the capstan problem, such as those of Groseberg and Plate \cite{grosbergplate1968}, McGee \cite{mcgee1977}, Beflosky \cite{beflosky1973}, Jung et al. \cite{jung2004,jung2008}, and Gao et al. \cite{gaowanghao2015}, have also circumvented this issue by prescribing the contact angle $\phi_c$ instead of the wrap angle $\phi$.

Another critical consequence of incorporating bending elasticity of the filament in the capstan problem, also pointed out by Stuart in \cite{stuart1961}, is the possible occurrence of point reaction forces at the contact boundaries \cite{oreillyvaradi03,oreilly2017}.
Any such forces would induce jumps in the internal force of the rod, and may contribute significantly to the computation of the output load $F_L$.
While some authors, such as Stuart \cite{stuart1961}, Groseberg and Plate \cite{grosbergplate1968}, and McGee \cite{mcgee1977}, have considered point forces at the contact boundaries in their models, those considerations have largely been ad hoc.
Other authors, such as Beflosky \cite{beflosky1973,beflosky1976}, Jung et al. \cite{jung2004,jung2008}, and Gao et al. \cite{gaowanghao2015}, have altogether ignored the possibility of point reaction forces at the contact boundaries.

In this article, we generalize the classic capstan problem by treating the filament as an elastic rod of finite thickness.
We refer to this system as the \emph{generalized capstan problem}, where we study both frictionless and frictional contact between the rod and the capstan.
Contrary to the approaches taken in the literature on the problem, we treat the two ends (i.e., points $s_1$ and $s_2$ in Fig. \ref{fig:introduction_schematics}b) of the contact region  as \emph{free boundaries}, namely boundaries whose locations cannot be prescribed a priori, but must be determined as part of the solution \cite{burridgekeller1978,oreilly2017}.
A distinguishing ingredient of the solution presented in this article is a jump condition -- which we derive systematically using the principle of virtual work -- valid at the contact boundaries.
The said jump condition not only locates the contact boundaries on the rod's centerline (and on the capstan), but also determines the jumps that the internal force and the curvature of the rod may suffer across such points.

In the frictionless contact case, we compute three qualitatively distinct kinds of rod equilibria, i.e., where the rod touches the capstan at one point, along a continuous arc, and at two points (Fig. \ref{fig:three_contact_configs}).
Our treatment of the frictionless case reveals that an elastic rod of \emph{finite length} wrapped around a capstan does not require friction to support unequal loads at its two ends.
In other words, force amplification across a finite length of an elastic rod can be achieved entirely by virtue of its bending elasticity.
This is in stark contrast to the classic case where the absence of friction, i.e., $\mu=0$, implies $F_0=F_L$ for any wrap angle $\phi$ of the filament, as is evident from equation \eqref{eq:classic_capstan}.
We classify the three aforementioned kinds of  equilibria in a limit where the length of the rod is much larger than the radius of the capstan.
In the same limit, we incorporate frictional interaction between the capstan and the elastic rod, and obtain a generalization of equation \eqref{eq:classic_capstan}.
We show that the maximum force ratio $F_L/F_0$ delivered by the generalized solution is a function of both $\phi$ and $F_0$, as opposed to the classic case where the force ratio is independent of the latter (see equation \eqref{eq:classic_capstan}).

Generalizations of the capstan problem in different directions have also been treated in the literature. 
The effects of nonlinear friction laws on the capstan problem have been considered by Lodge and Howell \cite{lodgehowell1954}, Beflosky \cite{beflosky1976}, Jung et al. \cite{jung2008} and Gao et al. \cite{gaowanghao2015},
while Liu and Vaz \cite{liuvaz2016} have considered the capstan problem with external pressure  applied to the filament.
A notable generalization of the capstan equation to thin inextensible strings lying on arbitrary surfaces was done by Maddocks and Keller \cite{maddockskeller1987}, and was revisited later from a variational perspective by Konyukhov \cite{konyukhov2015}.
Grandgeorge et al. \cite{grandgeorge2021} performed an experimental and theoretical study of two filaments in tight orthogonal contact, which serves as a generalization of the classic capstan problem by accounting for the filament thickness, and replacing the rigid circular capstan altogether with an identical filament.
A system of a closed flexible belt hanging on two pulleys has been studied by Belyav et al. \cite{belyaev2017} and Vetyukov et al. \cite{vetyukov2019}, while transient dynamics of a belt-pulley system with dry friction has been analyzed by Oborin et al. \cite{oborin2018}.
More recently, Grandgeorge et al. \cite{grandgeorge2022} have presented a study on the stationary dynamics of a sliding elastic rod in frictional contact with a circular capstan, with applications to belt-driven pulley systems.

The structure of this article is as follows.
We begin in Sect. \ref{sec:overview_of_theory} with a brief overview of the standard Cosserat rod theory, where we outline the balance laws, jump conditions, conservation laws, and constitutive assumptions.
We setup our specific problem of interest in Sect. \ref{sec:problem_setup}, and obtain relevant equations for planar deformations of an elastic rod wrapped around a rigid circular capstan.
In Sect. \ref{sec:frictionless_contact_finite_length}, we compute three distinct kinds of frictionless contact equilibria for a \emph{finite length} of an elastic rod, where the rod touches the capstan at one point, along a continuous arc, and at two points connected by an intermediate contact-free (or lift-off) region of the rod.
We also compute configurations where the elastic rod in frictionless contact with the capstan supports unequal loads at its two ends.
In Sect. \ref{sec:frictionless_contact_infinite_length}, we consider a \emph{long length} limit where the length $L$ of the rod is assumed much larger than the radius $R$ of the capstan (i.e. $L/R\gg 1$), and classify the three kinds of rod equilibria from Sect. \ref{sec:frictionless_contact_finite_length} in this limit.
Thereafter, in Sect. \ref{sec:frictional_contact}, we introduce frictional interaction between the capstan and the rod in the long length limit, and obtain a generalization of equation \eqref{eq:classic_capstan}.
Finally, we end with conclusions in Sect. \ref{sec:conclusion}.

\section{Overview of the essential theory}\label{sec:overview_of_theory}
Following Antman \cite{antman05}, we identify any configuration of an elastic rod with its centerline curve $\bx(s)\in\mathbb{R}^3$, and an ordered orthonormal frame of directors $\bd_i(s)$, $i \in\{1,2,3\}$, attached to it.
The parameter $s$ is the arc-length coordinate of the centerline in some reference configuration.
We will restrict our discussion to rods that are inextensible and unshearable, so that $s$ remains the arc-length coordinate in any configuration.
The inextensibility and unshearablity constraint, along with the orthonormality of the director frame, can be expressed, respectively, by the relations,
\begin{align}
\bx' = \bd_3\, ,\qquad\bd_i' = \bu\times\bd_i\,.\label{eq:centreline_and_director_evolution}
\end{align}
Here $\bu(s)$ is the Darboux (or strain) vector associated with the director frame, and the prime denotes derivative w.r.t. $s$. 
The director components $u_i := \bu\cdot\bd_i$, $i \in\{1,2,3\}$, of the Darboux vector measure the bending strains of the rod about the directors $\bd_i$.

For a cross-section of the rod centered at $s$, we denote by $\bn(s)$ and $\bmo(s)$, respectively, the net internal force and internal moment exerted by the material in $s^+$ on the material in $s^-$, where $s^\pm = \lim_{\epsilon\to 0} (s \pm \epsilon)$, and $\epsilon>0$.
The force and moment balance of a fixed material segment $[s_1,s_2]$ are then given by,
\begin{subequations}\label{eq:force_and_moment_integral_balance}
\begin{align}
\left[\bn\right]_{s_1}^{s_2} + \int_{s_1}^{s_2}\!\!\bp\, ds &= \bO\, ,\label{eq:force_balance_integral}\\
\left[\bmo + \bx\times\bn\right]_{s_1}^{s_2} + \int_{s_1}^{s_2}\!\!\bx\times\bp\, ds + \int_{s_1}^{s_2}\!\!\bl\, ds& = \bO\, ,\label{eq:moment_balance_integral}
\end{align}
\end{subequations}
where $\bp(s)$ and $\bl(s)$ are the external force and moment densities (per-unit length) distributed over the material segment.
When all the fields appearing in \eqref{eq:force_and_moment_integral_balance} are sufficiently regular in $[s_1,s_2]$, the two integral balances can be localized to obtain the following pointwise force and moment balance equations,
\begin{subequations}\label{eq:force_and_moment_local_balance}
\begin{align}
\bn' + \bp &= \bO\, ,\label{eq:force_balance_local}\\
\bmo' + \bx'\times\bn + \bl &=\bO\, .\label{eq:moment_balance_local}
\end{align}
\end{subequations}

Let $s_0\in[s_1,s_2]$ be a point around which $\bp$ and $\bl$ are highly localized.
At such a point, we idealize the two fields using the description \cite{oreillyvaradi99},
\begin{align}
	\bp=\bP\,\delta(s-s_0)\, ,\qquad \bl=\bL\,\delta(s-s_0)\, ,\label{eq:force_moment_deltas}
\end{align}
where $\delta(s)$ is the standard Dirac-delta function.
We also require the position and tangent vectors to be continuous, i.e., $\jump{\bx} = \bm 0$ and $\jump{\bx'} = \bm 0$, where $\jump{A} = A^+ - A^-$ denotes the jump in $A$, across any such points.
Then, using \eqref{eq:force_moment_deltas}, we localize \eqref{eq:force_and_moment_integral_balance} around the singular point $s_0$ to obtain the following force and moment jump conditions \cite{oreillyvaradi99,oreilly2017},
\begin{subequations}\label{eq:force_and_moment_jump_conditions}
\begin{align}
\jump{\bn} + \bP &=\bO\, ,\label{eq:force_jump_condition}\\
\jump{\bmo} + \bL &=\bO\, .\label{eq:moment_jump_condition}
\end{align}
\end{subequations}
The source terms $\bP$ and $\bL$ are identified, respectively, as the point force and the point moment exerted on the rod at $s=s_0$ by its environment \footnote{For a discussion on the existence of singular forces in contact problems with various beam theories see \cite{naghdirubin1989}.}.

To obtain a closed system of equations for the rod in regions away from any singular points, we introduce constitutive relations by assuming the rod to be hyperelastic with a straight and uniform natural configuration.
This means that there exists a scalar valued strain energy density function $W(\bu)$ for the rod, such that,
\begin{align}
	\bmo = \frac{\partial W(\bu)}{\partial\bu}\, .\label{eq:constitutive_relation}
\end{align}
We further assume the following quadratic form for the energy function,
\begin{align}
	W(\bu) = \frac{1}{2}\bu\cdot\bB\bu =\sum_{i=1}^{3} \frac{1}{2}B_i u^2_i\, ,\label{eq:energy_function}
\end{align}
where $\bB$ is a $3\times 3$ diagonal positive-definite stiffness matrix of the rod.
Using \eqref{eq:energy_function} in \eqref{eq:constitutive_relation}, we obtain,
\begin{align}
	\bmo=\bB\bu\, ,\label{eq:linear constitutive}
\end{align}
which delivers the internal moment as a linear function of the Darboux vector.

\subsection{Conservation laws}
In regions where $\bn$ and $\bmo$ are sufficiently regular,  equations \eqref{eq:force_and_moment_local_balance} imply the following conservation laws in absence of external force density $\bp$ and moment density $\bl$,
\begin{subequations}\label{eq:force_and_moment_conservation_laws}
\begin{align}
	\bn(s)& = \bn(0)\, ,\label{eq:force_integral}\\
	\bmo(s) + \bx(s)\times\bn(s) &= \bmo(0) + \bx(0)\times\bn(0)\, ,\label{eq:moment_integral}
\end{align}
\end{subequations}
where \eqref{eq:force_integral} has been used to arrive at \eqref{eq:moment_integral}. 
These force and moment conservation laws are, respectively, consequences of the translational and rotational invariance of the rod in ambient space, and are insensitive to the geometry and the material constituents of the rod \cite{maddocksdichman1994}.
In the generalised capstan problem, these conservation laws will hold only in the contact-free regions of the rod.

Next, we define the following function on the centerline of the elastic rod, which will prove to be of great significance in the upcoming analysis,
\begin{align}
H(s) := \bn\cdot\bx' + \bmo\cdot\bu - W(\bu)\, .\label{eq:Hamiltonian_definition}
\end{align}
We will refer to this function as the \emph{Hamiltonian function}\footnote{Our choice of the term ``Hamiltonian'' in this context is admittedly a slight abuse of terminology. A Hamiltonian, by definition, is an explicit function of the phase space variables of a dynamical system. We, however, treat it as an implicit function of the arc-length coordinate $s$, whose numerical value for any given configuration of the rod coincides with the proper Hamiltonian function for hyperelastic rods defined by Dichmann and Maddocks \cite{dichmannmaddocks1996}. Other workers (including the present author) have used different names, such as ``contact material force'' \cite{oreilly07} and ``material stress/force'' \cite{singhhanna2017elastica,hannasingh2018}, in the past to refer to the function defined in \eqref{eq:Hamiltonian_definition}.}.
Differentiating \eqref{eq:Hamiltonian_definition} w.r.t. $s$, and using \eqref{eq:force_and_moment_local_balance} and \eqref{eq:centreline_and_director_evolution}, it can be shown that in the absence of $\bp$ and $\bl$, the Hamiltonian function \eqref{eq:Hamiltonian_definition} is conserved along $s$ \cite{love2013,ericksen1970,antman1974,oreilly07,maddocksdichman1994,dichmannmaddocks1996,oreilly2017,nizettegoriely99,singhhanna2017elastica}, i.e.,
\begin{align}
	H(s)=H(0)\, .\label{eq:Hamiltonian_conservation}
\end{align} 
This law results from the translational symmetry of the rod in the arc-length coordinate $s$, and is valid only for hyperelastic constitutive relations \cite{steigmannfaulkner1993,maddocksdichman1994}.

As we will see later, unlike the force and moment conservation laws stated in \eqref{eq:force_and_moment_conservation_laws},  the conservation of the Hamiltonian function \eqref{eq:Hamiltonian_conservation} may persist even in the presence of external load densities, particularly when the rod is in frictionless contact with a rigid surface.
We will prove and make extensive use of this fact in the upcoming analysis.

For isotropic rods with no external moment density $\bl$, the twist $\bmo\cdot\bd_3$ is another conserved quantity.
However, for planar deformations (as in the generalized capstan problem), this conservation law holds identically true as a result of the moment balance \eqref{eq:moment_balance_local}, and offers no further utility.
We will therefore not expound upon this law any further.

\subsection{Jump condition at free boundaries}\label{subsec:matching_conditions}
A free boundary is a boundary whose location in the material is unknown a priori, and must be computed as part of the solution \cite{burridgekeller1978}.
A propagating crack in a solid, and a shock wave in a fluid, are examples of such boundaries.
In the present context of an elastic rod wrapped around a capstan, the boundaries of a contact region (such as $s_1$ and $s_2$ in Fig. \ref{fig:three_contact_configs}b), or any isolated points of contact (such as $s_1$ in Fig. \ref{fig:three_contact_configs}a, and $s_1$ and $s_2$ in Fig. \ref{fig:three_contact_configs}c) qualify as free boundaries.
In this subsection, we use the principle of virtual work to postulate a jump condition valid at such points.
The said jump condition will enable us to locate the free boundaries, and determine the jumps that the internal force and the curvature of the rod may suffer across such points.

Consider the following energy functional, augmented with the inextensibility and unshearability constraint, for a material segment $[s_1,s_2]$ of the rod,
\begin{align}
	\mathcal{E} = \int_{s_1}^{s_2}\!\!\left[ W(\bu) - \bn\cdot\left(\bx' - \bd_3\right)\right]ds\, ,\label{eq:energy}
\end{align}
We consider a transformation of the arc-length coordinate $s$ to $s^*$, where the latter is given by,
\begin{align}
s^*\equiv s+\delta s\, .\label{eq:arc_length_variation}
\end{align}
We stipulate that, under a transformation of the form $s\rightarrow s^*$,  the variation of functional \eqref{eq:energy} equals the net virtual work expended by the external force and moment densities, i.e.,
\begin{align}
\delta\mathcal{E} = \delta W_{ext}\, ,\label{eq:virtual_work_statement}
\end{align}
where,
\begin{align}
\delta W_{ext} = \int_{s_1}^{s_2}\!\!\bp\cdot\tilde\delta\bx\, ds + \int_{s_1}^{s_2}\!\!\bl\cdot\tdelta\bz\, ds\, .\label{eq:work}
\end{align}
Here $\tilde\delta\bx$ and $\tdelta\bz$ are first order variations\footnote{The variation $\tilde\delta$ is defined as $\tilde\delta\bx \equiv \bx^*(s) - \bx(s)$, such that the vector $\bx^*(s^*)$ and $\bx(s)$ correspond to the same material point. For more details see Appendix \ref{app:contact_condition_derivation}.} in the position vector, and an angle variable $\bz$ conjugate to $\bl$.

We assume that the material segment encapsulates a free boundary identified with an unknown arc-length coordinate $s_0\in[s_1,s_2]$, at which $\bp$ and $\bl$ admit the description given by \eqref{eq:force_moment_deltas}.
Computing $\delta\mathcal{E}$ due to $s\rightarrow s^*$, and localising \eqref{eq:virtual_work_statement} around $s=s_0$ leads to the following condition for configurations satisfying \eqref{eq:force_and_moment_local_balance} away from $s_0$,
\begin{align}
-\jump{\bn\cdot\bx' + \bmo\cdot\bu - W}\,\delta s_0 = \bP\cdot\tilde\delta\bx(s_0) + \bL\cdot\tdelta\bz(s_0)\, .\label{eq:virtual_work_localised}
\end{align}
For brevity of exposition, the details of the computations resulting in the above expression have been deferred to Appendix \ref{app:contact_condition_derivation}.
One can immediately recognise the expression in the double brackets in \eqref{eq:virtual_work_localised} as the Hamiltonian function $H$ defined in \eqref{eq:Hamiltonian_definition}, while the expression on the right can be interpreted as the net virtual work expended by the reaction force $\bP$ and moment $\bL$.

Next, we place a constitutive assumption on the mechanics of the free boundary:
We assume that the net work expended by the point force and the point moment at $s=s_0$, i.e., the right side of equation \eqref{eq:virtual_work_localised}, is zero\footnote{Equation \eqref{eq:virtual_work_localised} indicates that the net virtual work done by the point force and moment at $s=s_0$ must be absorbed by the internal structure of the interface at that point. Assuming this work to be zero is tantamount to assuming that the interface has no such internal structure.}. This reduces equation \eqref{eq:virtual_work_localised} to,
\begin{align}
-\jump{\bn\cdot\bx' + \bmo\cdot\bu - W}\,\delta s_0 = 0\, . \label{eq:virtual_work_equated_to_zero}
\end{align}
Since free boundaries by definition cannot be prescribed a priori, we have $\delta s_0\ne 0$, and consequently we conclude from above that the following jump condition must hold at $s=s_0$,
\begin{align}
\jump{H} = 0\, .\label{eq:Hamiltonian_jump_condition}
\end{align}
We will refer to this jump condition\footnote{Some authors also refer to \eqref{eq:Hamiltonian_jump_condition} as the Weierstrass-Erdmann corner condition, a term borrowed from the calculus of variations. However, we must point out that traditional derivations of these corner conditions \cite{gelfandfomin2000,bolza1973} do not consider Dirac-delta functions in the description of the integrand of the functional. Consequently, these derivations arrive directly at expressions equivalent to \eqref{eq:Hamiltonian_jump_condition}, skipping the intermediate step \eqref{eq:virtual_work_localised}.}
 as the \emph{free boundary condition} from here onward.
 
 There are other incarnations of \eqref{eq:Hamiltonian_jump_condition} present in the literature on rod mechanics. 
 For instance, one can obtain \eqref{eq:Hamiltonian_jump_condition} from the static version of the balance of ``material momentum'' for rods, posited by O'Reilly, by identifying $\mathsf{C} = -H$ and $\mathsf{B} = 0$ in equation (23) of \cite{oreilly07}.
 For an alternative approach to free boundary problems in rods using a quasi-static balance of energy, in lieu of \eqref{eq:Hamiltonian_jump_condition}, the reader may refer to \cite{hannasingh2018}.

\section{Problem setup}\label{sec:problem_setup}
We now specialize the theory presented so far to planar deformations, and obtain relevant equations for the contact and contact-free regions of an elastic rod wrapped around a circular capstan.

The internal force, internal moment, and the strain vector associated with a planar configuration can be given the following representations,
\begin{align}
	\bn = n_1\bd_1 + n_3\bd_3\, ,\qquad  \bmo = m_2\bd_2\, ,\qquad \bu=u_2\bd_2\, .\label{eq:planar_force_moment}
\end{align}
Similarly, equations \eqref{eq:linear constitutive} and \eqref{eq:Hamiltonian_definition} can written for planar deformations using \eqref{eq:centreline_and_director_evolution}$_1$, \eqref{eq:energy_function}, and \eqref{eq:planar_force_moment}, as,
\begin{align}
	m_2=B_2u_2\, ,\qquad H = n_3 + \frac{m_2^2}{2B_2}\, ,\label{eq:planar_constitutive_Hamiltonian}
\end{align}
where $B_2$ is the second diagonal component of the stiffness matrix $\bB$ of the rod, and represents its bending modulus about $\bd_2$.

\subsection{Contact region}
\begin{figure}[h!]
	\centering
		\includegraphics[width=0.95\linewidth]{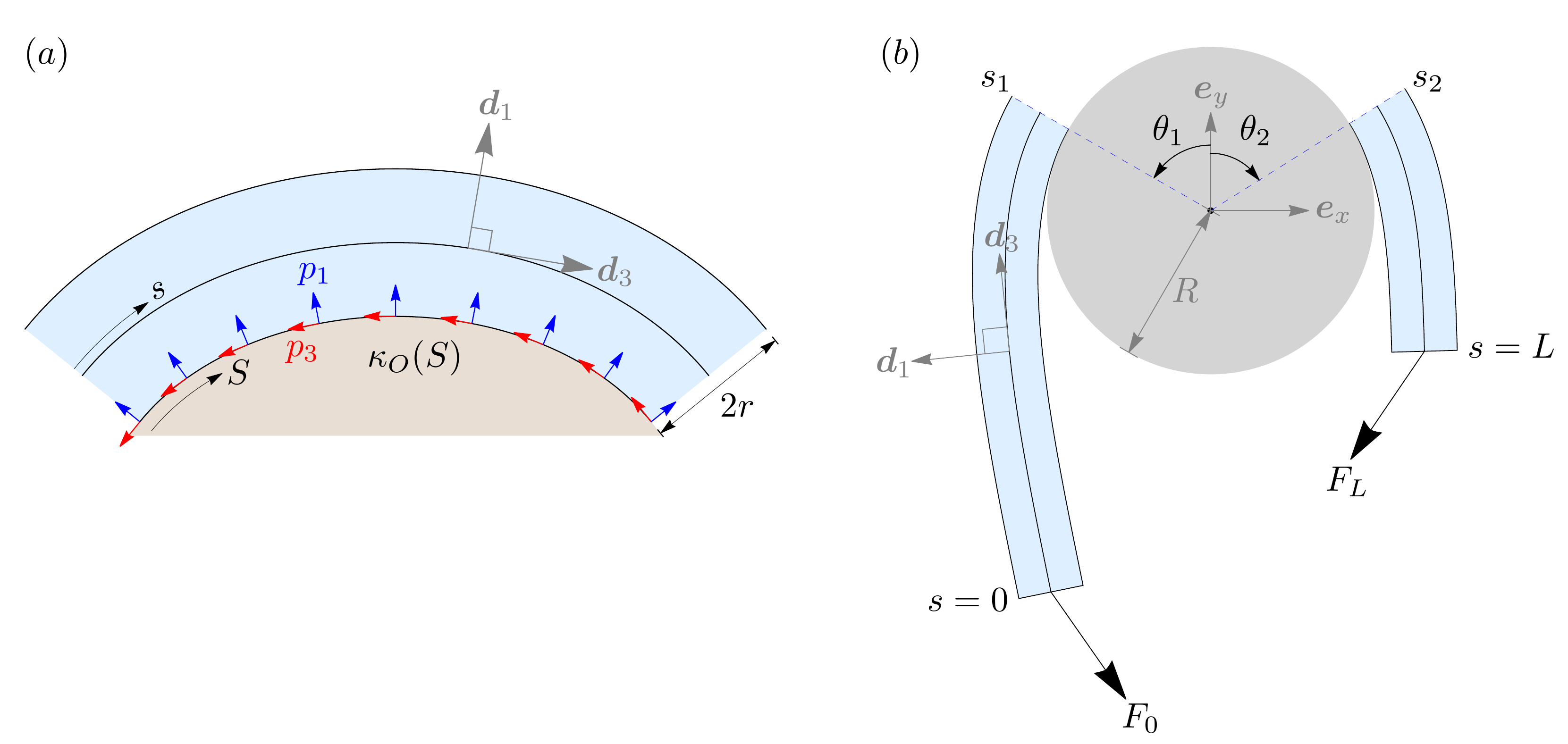}
	\caption{(a) An elastic rod in continuous planar contact with an obstacle with convex boundary of \emph{variable} curvature $\kappa_O(S)$. Here $S$ is a coordinate along the boundary of the obstacle, and is in one-to-one correspondence with $s$ along the centerline of the rod. (b) Contact-free tails of an elastic rod wrapped around a \emph{circular} capstan. The incoming and outgoing tails make contact with the capstan at $s=s_1$ and $s=s_2$ respectively, while the corresponding points on the capstan are identified with angles $\theta_1$ and $\theta_2$. Angles are defined positive when measured counter-clockwise.  }
	\label{fig:capstan_contact_and_tails_schematics}
\end{figure}

Consider a thick elastic rod in continuous planar contact with a rigid obstacle with a convex boundary, as shown in Fig. \ref{fig:capstan_contact_and_tails_schematics}a.
For simplicity, we will assume the rod to be of circular cross-section of radius $r$.
The force and moment densities exerted by the obstacle on the rod can then be represented in the director basis as,
\begin{subequations}\label{eq:force_and_moment_density_representation}
\begin{align}
\bp &= p_1\bd_1 - p_3\bd_3\, ,\qquad p_1\ge 0\,,p_3\ge 0\, ,\label{eq:force_density_representation}\\
\bl &= - r\bd_1\times \bp\, ,\nonumber \\
&= -r p_3 \bd_2\, ,\label{eq:moment_density_representation}
\end{align}
\end{subequations}
where we have assumed, without loosing generality, that the frictional component $p_3$ acts against the direction of the arc-length parametrization of the centerline.

Using \eqref{eq:force_and_moment_density_representation}, \eqref{eq:planar_force_moment} and \eqref{eq:centreline_and_director_evolution}$_2$, the $\bd_1$ and $\bd_3$ components of the force balance \eqref{eq:force_balance_local}, and the $\bd_2$ component of the moment balance \eqref{eq:moment_balance_local}, can be, respectively, written as,
\begin{subequations}\label{eq:force_and_moment_balance_components}
\begin{align}
	n_3' - n_1 u_2 - p_3 &= 0\, ,\label{eq:force_balance_d3}\\
	n_1' + n_3 u_2 + p_1 & = 0\, ,\label{eq:force_balance_d1}\\
	n_1 -  r p_3 + B_2 u'_2 & = 0\, .\label{eq:moment_balance_planar}
\end{align}
\end{subequations}
Furthermore, we identify the bending strain, given by $u_2$ for planar deformations, with the Frenet curvature $\kappa>0$ as,
\begin{align}
	u_2=-\kappa\, .\label{eq:frenet_curvature}
\end{align}

Using \eqref{eq:planar_constitutive_Hamiltonian} and \eqref{eq:moment_balance_planar}, we eliminate $n_3$ and $n_1$ in favour of $H$ and $p_3$ in \eqref{eq:force_balance_d3} and \eqref{eq:force_balance_d1}, and non-dimensionalize the variables as $r\rightarrow r/\tilde{R}$, $\kappa\rightarrow\tilde{R}\kappa$, $H\rightarrow H/\tilde{F}$, $p_1\rightarrow(\tilde{R}/\tilde{F})p_1$, $p_3\rightarrow(\tilde{R}/\tilde{F})p_3$ using some characteristic length scale $\tilde{R}$ and force scale $\tilde{F}$.
The result is the following dimensionless set of equations,
\begin{subequations}\label{eq:Hamiltonian_and_friction_ND}
\begin{align}
H' &= p_3(1- r \kappa)\, ,\label{eq:Hamiltonian_contact_ND}\\
r p_3' &= H \kappa - p_1 - B\left(\frac{1}{2}\kappa^3 + \kappa''\right)\, ,\label{eq:friction_contact_ND}
\end{align}
\end{subequations}
where $B = B_2/(\tilde{F}\tilde{R}^2)$ is the dimensionless bending stiffness of the rod.
In the contact region, the centerline of the rod and the boundary of the obstacle form a pair of parallel (or offset) curves, separated uniformly by distance $r$. As a consequence, the curvature $\kappa(s)$ of the centerline is related to the curvature $\kappa_O(S)$ of the obstacle boundary as \cite{farouki1990}
 \begin{align}
 \kappa(s) = \frac{\kappa_O(S)}{1+r\kappa_O(S)}\, .\label{eq:curvatures_relation}
\end{align} 
Here $S$ is the arc-length coordinate along the obstacle boundary, and is in some one-to-one correspondence with $s$.
Given this correspondence, along with $\kappa_O(S)$ for a given obstacle,  $\kappa(s)$ becomes a known function.\footnote{Relation \eqref{eq:curvatures_relation} can be obtained by invoking a geometric property of parallel curves, namely that their radii of curvatures must differ by $r$ for all $s$, i.e. $\kappa(s)^{-1} -\kappa_O(S)^{-1}=r$. This expression delivers \eqref{eq:curvatures_relation} on rearrangement.}

Equations \eqref{eq:Hamiltonian_and_friction_ND} can also be used in the contact-free regions by substituting $p_1=p_3=0$, and treating $\kappa(s)$ as an unknown function to be determined by integrating \eqref{eq:friction_contact_ND} under an appropriate set of boundary conditions.

\subsection{Contact-free region}\label{sec:circular_capstan_problem_setup}
Consider Fig. \ref{fig:capstan_contact_and_tails_schematics}b, which shows the schematic of two contact-free tails of a terminally loaded elastic rod wrapped around a circular capstan of radius $R$.
We identify this radius as the characteristic length scale of the problem, and therefore, from here onward, all lengths will be stated in units of $R$.
We will refer to the two contact-free regions touching the left and the right half of the capstan as the \emph{incoming} and \emph{outgoing} tail respectively.
Here we derive formulas needed to determine the location of the contact points, labeled by $\theta_1$ and $\theta_2$ on the capstan, and $s_1$ and $s_2$ on the rod's centerline.

We assume a Cartesian frame $\{\be_x,\be_y\}$ such that its origin coincides with the center of the capstan, and $\be_y$ bisects the angle $\pi-\phi$ (see Fig. \ref{fig:introduction_schematics}).
The two end forces can then be accorded the representations $\bF_0 = -F_0\left[\cos(\phi/2)\be_x + \sin(\phi/2)\be_y\right] $ and $\bF_L = F_L\left[\cos(\phi/2)\be_x -\sin(\phi/2)\be_y\right]$.
Similarly, the tangents at the two points of contact can be represented as $\bd_3(s_1) = \cos\theta_1\be_x + \sin\theta_1\be_y$ and $\bd_3(s_2) = \cos\theta_2\be_x + \sin\theta_2\be_y$, where $\theta_1$ and $\theta_2$ are angles as defined in Fig. \ref{fig:capstan_contact_and_tails_schematics}b.
Using these representations, together with the conservation of the internal force \eqref{eq:force_integral},  and the conservation of the Hamiltonian function \eqref{eq:Hamiltonian_conservation}, $\theta_1$ and $\theta_2$ can be written as,
\begin{align}
	\theta_1 = \frac{\phi}{2} - \arccos\left[\frac{1}{F_0}\left(H_0-\frac{1}{2}B \kappa_1^2\right)\right]\, ,\qquad\theta_2 = -\frac{\phi}{2} + \arccos\left[\frac{1}{F_L}\left(H_L-\frac{1}{2}B\kappa_2^2\right)\right]\, .\label{eq:angles_unknown}
\end{align}
Here $\kappa_1=\kappa(s_1^-)$ and $\kappa_2=\kappa(s_2^+)$, and $H_0$ and $H_L$ are the constant numerical values of the Hamiltonian function in the incoming and outgoing tails respectively.

 To locate the points $s_1$ and $s_2$ on the centerline of the rod, we use \eqref{eq:Hamiltonian_and_friction_ND} for the incoming and outgoing tails with $p_1=p_3=0$.
Equation \eqref{eq:Hamiltonian_contact_ND} then simply reduces to $H'=0$ confirming the conservation of the Hamiltonian function in the two tails, while \eqref{eq:friction_contact_ND} reduces to $0=H\kappa - B\left(\tfrac{1}{2}\kappa^3 + \kappa''\right)$.\footnote{This is the planar version of a more general equation derived by Langer and Singer \cite{langersinger1996} governing three dimensional deformations of Kirchhoff elastic rods. }
Integrating this equation after multiplying it with $2B\kappa'$, and identifying the constant of integration as $|\bn|^2 - H^2$ \cite{singhhanna2017elastica}, we obtain after some rearrangement,
\begin{align}
	\left(B\kappa'\right)^2 + \left(\frac{1}{2}B\kappa^2 - H\right)^2 = |\bn|^2\, .\label{eq:curvature_equation}
\end{align} 
Upon resolving the above equation for the positive root of $\kappa'$ and integrating it again, we obtain the following expression for the length $l_{AB}$ of the rod between any two points $s_A$ and $s_B$ with curvatures $\kappa_A$ and $\kappa_B$,
\begin{align}
	l_{AB}=\int_{\kappa_A}^{\kappa_B}\frac{d\kappa}{\sqrt{\left(\frac{|\bn|-H}{B}+\tfrac{1}{2}\kappa^2 \right)\left(\frac{|\bn|+H}{B} -\tfrac{1}{2}\kappa^2\right)}} = \mathscr{L}\left(|\bn|,H;\kappa\right)\big\rvert_{\kappa_A}^{\kappa_B}\, .\label{eq:length_function}
\end{align}
The explicit expression for $\mathscr{L}$ in terms of Elliptic integral of the first kind is provided in Appendix \ref{app:elliptic_integrals}.
The arc-length coordinates $s_1$ and $s_2$ are then easily expressed using \eqref{eq:length_function} between the two ends of the incoming and outgoing tails,

\begin{align}
	s_1 =\mathscr{L}(F_0, H_0;\kappa)|_{0}^{\kappa_1}\qquad s_2 = L-\mathscr{L}(F_L, H_L;\kappa)|_0^{\kappa_2}\, .\label{eq:arc_lengths_coordinates}
\end{align}
Once $\theta_1$ and $\theta_2$ along with $s_1$ and $s_2$ are determined, the incoming and outgoing tails can be computed by integrating the following set of equations,
\begin{subequations}\label{eq:configuration}
\begin{align}
\theta' &= -\kappa\, ,\qquad x' = \cos\theta\, ,\qquad y' =\sin\theta\, ,\label{eq:configuration_kinematics}\\
B\kappa' &= n_1\, ,\qquad n_1' =n_3\kappa\, ,\qquad n_3'=-n_1\kappa\, .\label{eq:configuration_mechanics}
\end{align}
\end{subequations}
Equations in \eqref{eq:configuration_kinematics} are obtained using \eqref{eq:planar_force_moment}$_3$, \eqref{eq:frenet_curvature}, and the representations $\bx=x\be_x+y\be_y$ and $\bd_3(s) = \cos\theta \be_x + \sin\theta\be_y$ in \eqref{eq:centreline_and_director_evolution}. 
While  \eqref{eq:configuration_mechanics} are obtained from \eqref{eq:force_and_moment_balance_components} using $p_1=p_3=0$ and \eqref{eq:frenet_curvature}.
For the incoming tail, equations \eqref{eq:configuration} can be integrated from $s_1$ to $0$ using the following initial conditions,
\begin{subequations}\label{eq:incoming_initial_conditions}
	\begin{align}
		&\theta(s_1^-) = \theta_1\, ,\quad x(s_1^-) = -(1+r)\sin\theta_1\, ,\quad y(s_1^-) = (1+r)\cos\theta_1\, ,\\
		&\kappa(s_1^-) = \kappa_1\, ,\quad n_1(s_1^-) = F_0\sin\left(\frac{\phi}{2}-\theta_1\right)\, ,\quad n_3(s_1^-) = F_0\cos\left(\frac{\phi}{2}-\theta_1\right)\,,
	\end{align}
\end{subequations}
whereas the configuration for the outgoing tail can be obtained by integrating \eqref{eq:configuration} from $s_2$ to $L$ using the following initial conditions,
\begin{subequations}\label{eq:outgoing_initial_conditions}
	\begin{align}
		&\theta(s_2^+) = \theta_2\, ,\quad x(s_2^+) = -(1+r)\sin\theta_2\, ,\quad y(s_2^+) = (1+r)\cos\theta_2\, ,\\
		&\kappa(s_2^+) = \kappa_2\, ,\quad n_1(s_2^+) = -F_L\sin\left(\frac{\phi}{2}+\theta_2\right)\, ,\quad n_3(s_2^+) = F_L\cos\left(\frac{\phi}{2}+\theta_2\right)\,.
	\end{align}
\end{subequations}
Note that, for simplicity, we will ignore any possible contact that may arise between the incoming and outgoing tails (such as in Fig. \ref{fig:three_contact_configs}c), i.e., the two tails will be allowed to pass through one another without physical interaction.

\subsection{Corollary of the free boundary condition}
Analogous to \eqref{eq:force_and_moment_density_representation}, a point reaction force at a free boundary can be written in the director frame as $\bP = P_1\bd_1 - P_3\bd_3$ and $\bL=-r P_3\bd_2$, with $P_1\ge 0, P_3\ge 0$.
Using these expressions, along with \eqref{eq:planar_force_moment} and \eqref{eq:planar_constitutive_Hamiltonian}, the director components of the force and moment jump conditions \eqref{eq:force_and_moment_jump_conditions} are written as,
\begin{align}
\jump{n_3} - P_3 = 0\, ,\qquad\jump{r p_3 + B \kappa'} + P_1 = 0\, ,\qquad \jump{B\kappa} +  rP_3 = 0\, .\label{eq:jump_conditions_director_components}
\end{align}
Similarly, using \eqref{eq:planar_constitutive_Hamiltonian}, equation \eqref{eq:Hamiltonian_jump_condition} can be written as $\jump{n_3 + \tfrac{1}{2}B\kappa^2} = 0$, which upon using the algebraic identity $\jump{A^2} = 2\jump{A}\avg{A}$ and equation \eqref{eq:jump_conditions_director_components}$_3$ can be rearranged as,
\begin{align}
P_3\left(1- r \avg{\kappa}\right) = 0\, .\label{eq:Hamiltonian_rewritten}
\end{align}
Assuming that the term inside the bracket is non-zero\footnote{For a cylindrical tube of constant radius to remain smooth, the curvature function of its centerline must satisfy $\kappa<1/r$ \cite{gonzalezmaddocks1998}. Therefore, as a consequence we have $1-r\avg{\kappa}> 0$. }, we conclude from above that $P_3 = 0$.
Combining this observation with \eqref{eq:jump_conditions_director_components}, we conclude,
\begin{align}
\jump{n_3} = 0\, ,\qquad \jump{B\kappa} = 0\, ,\label{eq:tension_curvature_jump}
\end{align}
meaning that the tension and the curvature (and as a result the internal moment) of the elastic rod must be continuous across the free boundaries.
The rod may still, however, experience a jump in the shear component $n_1=rp_3+B\kappa'$, induced by a non-zero $P_1$ in \eqref{eq:jump_conditions_director_components}$_2$.

\section{Frictionless contact with finite length}\label{sec:frictionless_contact_finite_length}
We first consider frictionless contact of a \emph{finite} length $L$ of an elastic rod wrapped around a circular capstan.
We identify the following three qualitatively distinct kinds of equilibria:
1.) \emph{one-point contact}, where the rod touches the capstan at precisely one point (Fig. \ref{fig:three_contact_configs}a), 2.) \emph{line contact}, where the rod touches the capstan along a continuous arc (Fig. \ref{fig:three_contact_configs}b), and 3.) \emph{two-point contact}, where the rod touches the capstan at two distinct points, with the intermediate contact-free  length of the rod referred to as the \emph{lift-off} region (Fig. \ref{fig:three_contact_configs}c). 

A key property of the three kinds of frictionless equilibria stated above is that the Hamiltonian function $H(s)$ remains conserved throughout the length of the rod with the same numerical value, i.e.,
\begin{align}
	H(s) = H(0) = H \qquad \text{for}\qquad s\in[0,L].\label{eq:frictionless_Hamiltonian}
\end{align}
In other words, the conservation of the Hamiltonian function is \emph{insensitive} to  frictionless contact between the rod and the capstan.
Note that this fact does not depend on the shape of the capstan.
For \emph{one-point} and \emph{two-point} contact equilibria, statement \eqref{eq:frictionless_Hamiltonian} follows directly from \eqref{eq:Hamiltonian_conservation} and \eqref{eq:Hamiltonian_jump_condition}.
For \emph{line contact}, however, one must additionally show the conservation of $H$ in the contact region, which can be easily deduced from equation \eqref{eq:Hamiltonian_contact_ND} by substituting $p_3=0$.

\subsection{One-point contact}\label{sec:one_point_finite_length}
One-point contact equilibria entails the rod touching the capstan at exactly one point, where the centerline curvature is smaller than the critical value $\kappa_c = (1+r)^{-1}$, obtained from \eqref{eq:curvatures_relation} using $\kappa_O=1$ in units of $R$.
The two key unknowns needed to compute such a configuration are the location of the point of contact, and the rod's centerline curvature at that point.
To determine these two quantities, we will need the following two relations,
\begin{subequations}\label{eq:point_contact_relations}
\begin{align}
	\mathscr{L}(F_0,H;\kappa)|_{0}^{\kappa_1} + \mathscr{L}(F_L,H;\kappa)|_{0}^{\kappa_1} = L\, ,\label{eq:point_contact_length_relation}\\
	 \arccos\left[\frac{1}{F_0}\left(H - \frac{1}{2}B\kappa_1^2\right)\right] + \arccos\left[\frac{1}{F_L}\left(H - \frac{1}{2}B\kappa_1^2\right)\right] = \phi\, .\label{eq:point_contact_angle_relation}
\end{align}
\end{subequations}
The first equation above enforces the total length $L$ of the rod, and has been obtained using \eqref{eq:length_function}.
The second equation is obtained by enforcing $\theta_1 = \theta_2$ using \eqref{eq:angles_unknown}, and substituting in it $H_0 = H_L = H$ due to \eqref{eq:frictionless_Hamiltonian}, and $\kappa_2=\kappa_1$ due to \eqref{eq:tension_curvature_jump}$_2$.
For prescribed values of $F_0$, $F_L$, $\phi$, and $L$, equations \eqref{eq:point_contact_relations} can be numerically solved to obtain $\kappa_1$ and $H$, which on back substitution in \eqref{eq:angles_unknown} and \eqref{eq:arc_lengths_coordinates} delivers $\theta_1$ and $s_1$, respectively.

The entire configuration can then be constructed by solving two initial value problems: one for the incoming tail, and the other for the outgoing.
The former can be obtained by integrating \eqref{eq:configuration} from $s_1$ to $0$ using initial conditions \eqref{eq:incoming_initial_conditions}, and the latter by integrating \eqref{eq:configuration} from $s_2=s_1$ to $L$, using initial conditions \eqref{eq:outgoing_initial_conditions}.

\subsection{Line contact}\label{sec:line_contact_finite_length}
\begin{figure}[t]
	\centering
	\includegraphics[width=0.95\linewidth]{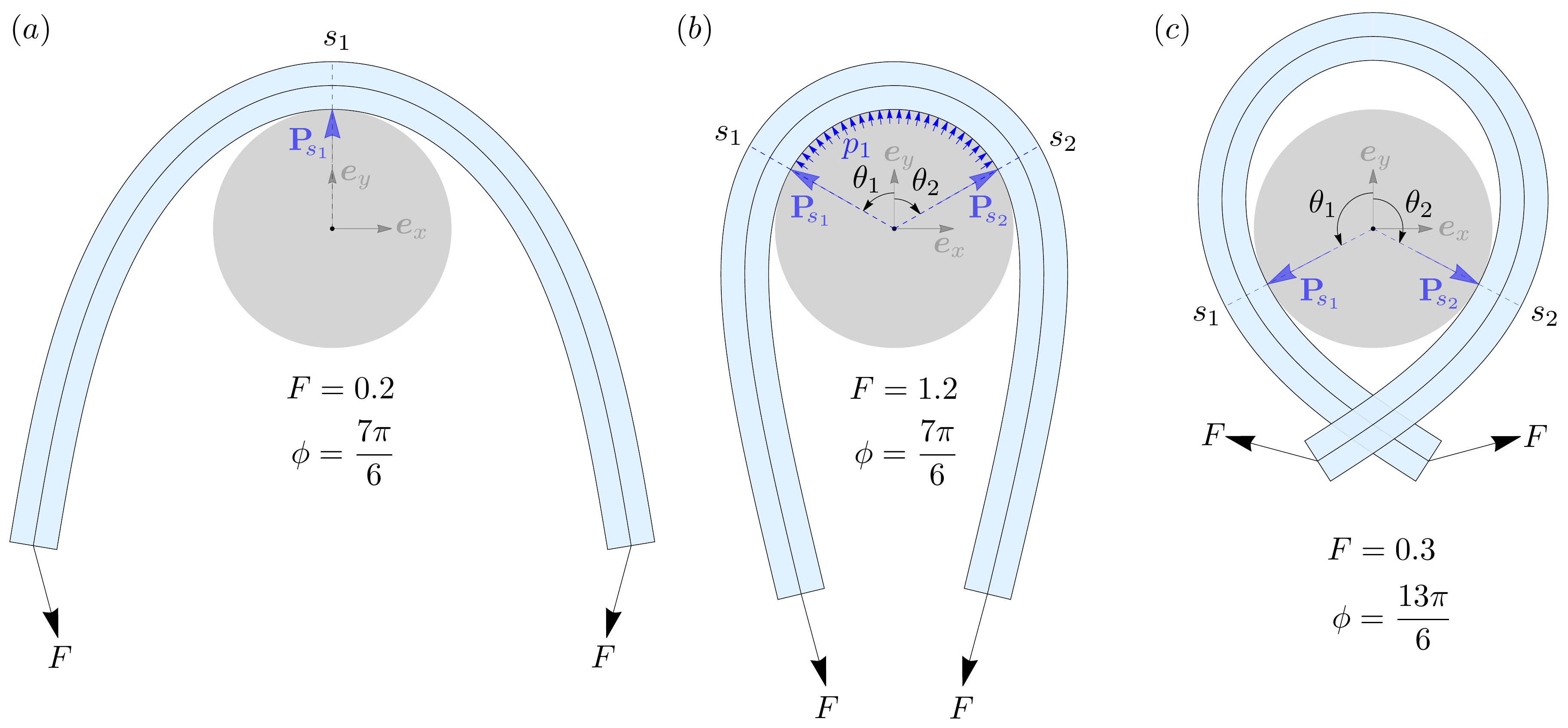}
	\caption{Three symmetric equilibrium configurations of an elastic rod in frictionless contact with a circular capstan. The rod has a length $L=10$, radius of cross-section $r=0.2$, and bending modulus $B=1$. (a) One-point contact with $\theta_1=\theta_2=0$ and $s_1=0.5L$. The capstan exerts a point reaction force $|\bP_{\!s_1}|=0.386$ in the normal direction at $s_1$. (b) Line contact with $s_1=0.374L$, $s_2=0.626L$, and $\theta_1=-\theta_2=1.052$ rad. The magnitude of the normal point reaction forces are $|\bP_{\!s_1}|=|\bP_{\!s_2}|=0.845$, and the distributed pressure density $p_1=0.71 $. (c) Two-point contact where $s_1=0.209L$, $s_2 = 0.791L$ and $\theta_1=-\theta_2=2.05$ rad. The magnitudes of the normal point reaction forces are $|\bP_{\!s_1}|=|\bP_{\!s_2}|=0.167$.}
	\label{fig:three_contact_configs}
\end{figure}

For configurations with line contact between the elastic rod and the capstan, the curvature of the rod's centerline in the contact region is identically equal to $\kappa_c=(1+r)^{-1}$.
Therefore, using \eqref{eq:tension_curvature_jump}$_2$ across the contact boundaries at $s_1$ and $s_2$ (Fig. \ref{fig:three_contact_configs}b), we conclude that $\kappa(s_1^-) = \kappa(s_2^+) = \kappa_c$.
With this observation, the only remaining unknown needed to compute the locations $\theta_1$ and $\theta_2$ from \eqref{eq:angles_unknown}, and $s_1$ and $s_2$ from \eqref{eq:arc_lengths_coordinates}, is $H_0=H_L=H$.
This can be computed by enforcing the total length of the rod using \eqref{eq:angles_unknown} and \eqref{eq:length_function} as,
\begin{align}
	\mathscr{L}(F_0,H;\kappa)|_{0}^{\kappa_c} + (1+r)\left\{\phi - \arccos\left[\frac{1}{F_0}\left(H - \frac{1}{2}B\kappa_c^2\right)\right]-\arccos\left[\frac{1}{F_L}\left(H - \frac{1}{2}B\kappa_c^2\right)\right]\right\} + \mathscr{L}(F_L,H;\kappa)|_{0}^{\kappa_c} = L\,, \label{eq:line_contact_relations}
\end{align}
where the middle term is simply $(1+r)(\theta_1-\theta_2)$ substituted with \eqref{eq:angles_unknown}, and measures the length of the centerline in the contact region.
 For prescribed values of $F_0$, $F_L$, $\phi$, and $L$, equation \eqref{eq:line_contact_relations} can be numerically solved for $H$, and  $\theta_1$, $\theta_2$, $s_1$, and $s_2$ can be computed by substituting $H$ back into \eqref{eq:angles_unknown} and \eqref{eq:arc_lengths_coordinates}.
Equations \eqref{eq:moment_balance_planar}, \eqref{eq:planar_constitutive_Hamiltonian}, and \eqref{eq:friction_contact_ND} then imply the following relations for $n_1$, $n_3$ and $p_1$ in the contact region, with $p_3=0$ and $\kappa =\kappa_c$,
\begin{align}
	n_1 = 0\, ,\qquad n_3 = H - \tfrac{1}{2}B\kappa_c^2\, ,\qquad p_1 = H\kappa_c - \frac{1}{2}B\kappa_c^3\, .\label{eq:line_contact_shaer_tension_pressure}
\end{align}

Finally, the incoming and outgoing tails can be computed by integrating \eqref{eq:configuration} using initial conditions \eqref{eq:incoming_initial_conditions} and \eqref{eq:outgoing_initial_conditions}, respectively.

\subsubsection{Force amplification with frictionless contact}\label{sec:force_amplification_frictionless}
It is worth emphasising here that for frictionless contact, the two kinds of equilibria described in subsections \ref{sec:one_point_finite_length} and \ref{sec:line_contact_finite_length} do not require the end loads $F_0$ and $F_L$ to be equal.
 The implication being that force amplification across an elastic rod wrapped around a circular capstan can be achieved without friction.
This is in stark contrast to the classic capstan problem governed by equation \eqref{eq:classic_capstan}, where the absence of friction, i.e., $\mu=0$, implies $F_0=F_L$.

\begin{figure}[t]
	\centering
	\includegraphics[width=0.9\linewidth]{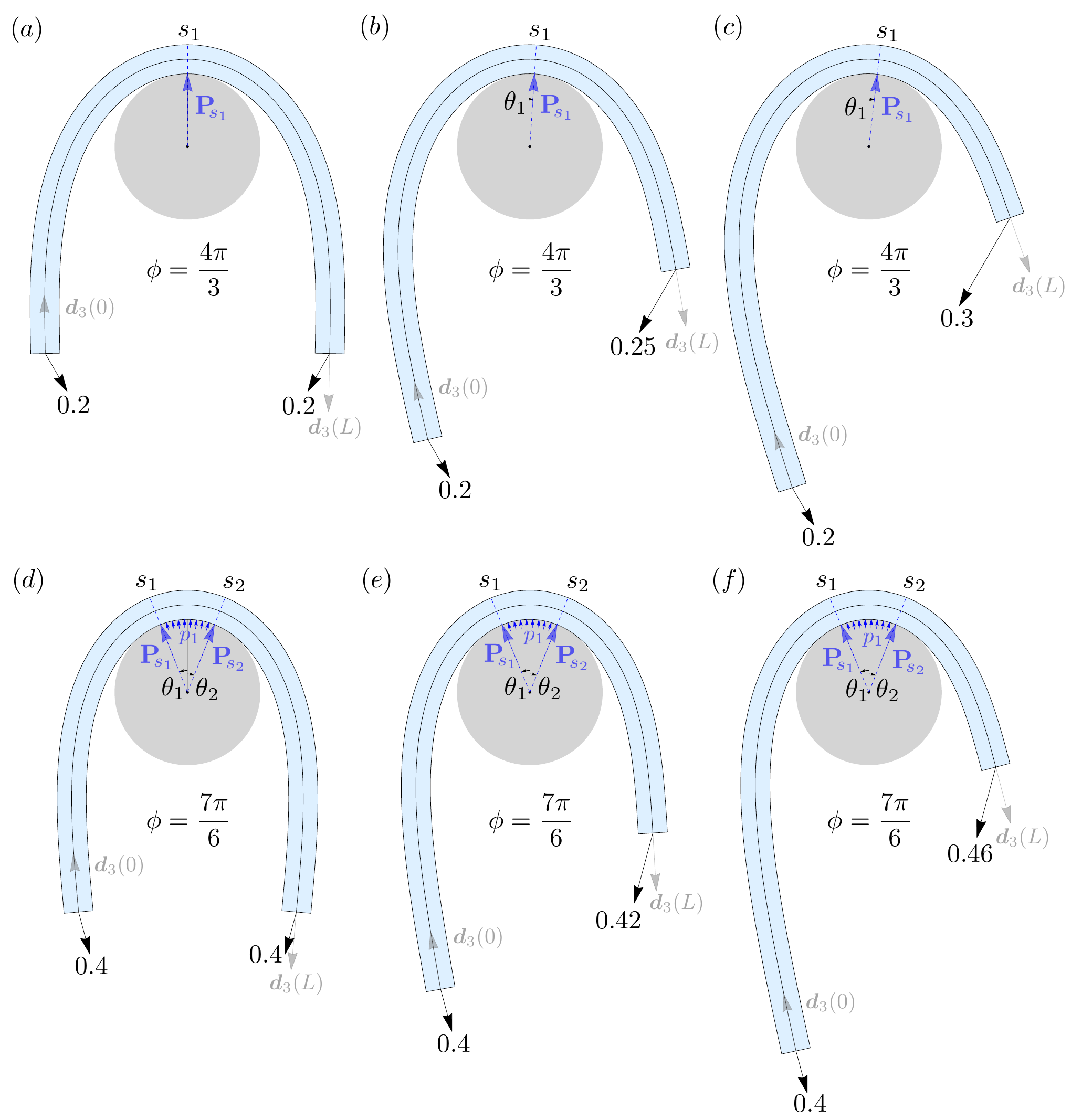}
	\caption{Some examples of asymmetric equilibria of an elastic rod with $r=0.2$, $L=10$, and $B=1$, in \emph{frictionless} contact with a circular capstan. Each row above depicts three configurations where the load on the left end, and the wrap angle $\phi$, are kept fixed, while the loads on the right end is increased. The numerical values $\{s_1,\theta_1,|\bP_{\!s_1}|\}$ in configurations (a), (b), and (c), respectively, are $\{5,0,0.346\}$, $\{6.25,-0.0641,0.391\}$,  $\{7.02,-0.115,0.436\}$. Similarly, the values of $\{s_1,s_2,\theta_1,\theta_2,|\bP_{\!s_1}|,|\bP_{\!s_2}|,p_1\}$ for configurations shown in (d), (e), and (f), respectively, are $\{4.54,5.46,0.379,-0.379,0.397,0.397,0.0391\}$, $\{5.61,6.54,0.39,-0.384,0.397,0.417,0.0427\}$, $\{6.47,7.4,0.393,-0.376,0.397,0.457,0.0436\}.$
	} 
	\label{fig:assymetric_equilibria}
\end{figure}

Configurations with unequal end loads for one-point and line contact equilibria are shown in Fig. \ref{fig:assymetric_equilibria}.
Although this ability of partially constrained elastic rods to support unequal loads may appear unintuitive at first, it is revealed quite remarkably by relation \eqref{eq:frictionless_Hamiltonian}.
Computing the Hamiltonian function at $s=0$ using \eqref{eq:planar_constitutive_Hamiltonian} as $H(0) = -F_0\cos\alpha_0$, and analogously $H(L)=F_L\cos\alpha_L$ ($\alpha_0$ and $\alpha_L$ being the smaller angles between the tangent and the force at $s=0$ and $s=L$, respectively),
we obtain using \eqref{eq:frictionless_Hamiltonian},
\begin{align}
	F_0\cos\alpha_0 + F_L\cos\alpha_L = 0\, .\label{eq:force_amplification_frictionless}
\end{align}
One can now clearly see that any inequality between $F_0$ and $F_L$ can be accommodated by the difference in the angles $\alpha_0$ and $\alpha_L$ (see Fig. \ref{fig:assymetric_equilibria}).
Therefore, in principle, for any given load $F_0$, an arbitrarily high load $F_L$ can be supported by the rod by assuming a configuration where $\alpha_0 \rightarrow \pi$ and $\alpha_L\rightarrow \pi/2$.

This property of a partially constrained elastic rod to be able to support unequal loads has been exploited in the design of the \emph{elastica arm scale} in \cite{bosi2014}\footnote{Equation \eqref{eq:force_amplification_frictionless} is equivalent to equation (2.11) of \cite{bosi2014}, which the authors of that article described as a ```geometrical condition' of equilibrium'' representing the balance of axial thrust on the rod. We also note that since \eqref{eq:frictionless_Hamiltonian} is independent of the shape of the constraining rigid body, its corollary \eqref{eq:force_amplification_frictionless} is equally applicable to problems involving frictionless sleeve constraints in \cite{bosi2014,bigoni15,bigoni14}.}, which is a weight measuring balance scale with deformable lever arms, as opposed to rigid arm scales that have been used to measure weight since time immemorial.
For an illuminating treatment of the elastica arm scale using a non-classical ``material force'' balance, the reader may refer to \cite{oreilly2015eshelby}.

 \subsection{Two-point contact with a lift off region}\label{sec:two_point_finite_length}
 If the wrap angle $\phi$ of the forces around the capstan is greater than $2\pi$, and the applied forces are small enough, the rod equilibria may transition from line contact to two-point contact, with an intermediate contact-free lift-off region (Fig. \ref{fig:three_contact_configs}c).
 Here we consider two-point equilibria which are symmetric under reflection across the $\be_y$ axis.
 Consequently, we will only compute equilibria for one half of the rod, while the other half is obtained by symmetry.
 We will also ignore any contact interactions between the incoming and outgoing tails of the rod.
  
 Consider a configuration as shown in Fig. \ref{fig:three_contact_configs}c, loaded with forces of magnitude $F$ with a wrap angle greater than $2\pi$. To compute such an equilibrium, we would need to determine the constant value $H$ of the Hamiltonian function, and the curvature $\kappa_1$ of the rod at the point of contact.
To this objective, we first write down the internal force $\bn_l$ in the lift-off region, and the curvature $\kappa_{L/2}$ of the rod at $s=L/2$, in terms of $H$ and $\kappa_1$ as follows,
\begin{align}
	\bn_l = \left(H-\frac{1}{2}B\kappa_1^2\right)\sec\theta_1\be_x\,,\qquad \kappa_{L/2} = \sqrt{\frac{2(H-n_l)}{B}}\, .\label{eq:internal_force_and_curvature_in_lift_off}
\end{align}
Here $\theta_1$ is given by \eqref{eq:angles_unknown}$_1$ with $F_0 = F$ and $H_0 = H$, and $n_l = |\bn_l|$ in the second expression above.
Equation \eqref{eq:internal_force_and_curvature_in_lift_off}$_1$ has been obtained using  $\bn_l = n_{l1}(s_1)\bd_1(s_1) + n_{l3}(s_1)\bd_3(s_1)$, requiring $\bn_l\cdot\be_y=0$ (by symmetry), and using $n_{l3}(s_1)=H-(1/2)B\kappa_1^2$.
Equation \eqref{eq:internal_force_and_curvature_in_lift_off}$_2$ has similarly been obtained using the fact that the tension at $s=L/2$ is given by $n_{l3}(L/2) = n_l = H-(1/2)B\kappa_{L/2}^2$.

 The two conditions needed to solve for $H$ and $\kappa_1$ can then be written as,
 \begin{subequations}\label{eq:two_point_contact_relations}
  \begin{align}
 	\mathscr{L}(F,H;\kappa)|_{0}^{\kappa_1} + \mathscr{L}(n_l,H;\kappa)|^{\kappa_{L/2}}_{\kappa_1} = \frac{L}{2}\, ,\label{eq:two_point_contact_length_relation}\\
 	\qquad -(1+r)\left(H - \frac{1}{2}B\kappa_1^2\right)\tan\theta_1 + \mathscr{P}\left(n_l,H;\kappa\right)\big\rvert_{\kappa_1}^{\kappa_{L/2}} = 0\, . \label{eq:two_point_contact_position_relation}
 \end{align}
\end{subequations}
 The first expression above, obtained using \eqref{eq:length_function}, is a statement enforcing the length of one symmetric half of the rod, 
 To obtain the second expression, we integrated the relation $n_{l3} = H-(1/2)B\kappa^2$ from $s_1$ to $L/2$, used the fact that $n_{l3} = \bn_l\cdot\bd_3 = \bn_l\cdot\bx' = (\bn_l\cdot\bx)'$, substituted $\bn_l\cdot\bx(s_1) = -(1+r)n_l\sin\theta_1$, and then finally enforced $\bn_l\cdot\bx(L/2) = 0$ (due to symmetry) in the resulting expression.
 The function $\mathscr{P}$ in the second expression results from a variable change in $\int_{s_1}^{L/2}(H-(1/2)B\kappa^2)ds$ from $s$ to $\kappa$ via $ds=d\kappa/\kappa'$, and using equation \eqref{eq:curvature_equation} to substitute for $\kappa'$ to yield,
 \begin{align}
 	\int_{\kappa_1}^{\kappa_{L/2}}\frac{\left( H - \tfrac{1}{2}B \kappa^2\right)}{\sqrt{\left( \frac{n_l - H}{B}+\tfrac{1}{2}\kappa^2\right)\left(\tfrac{n_l+H}{B} - \tfrac{1}{2}\kappa^2\right)}}\, d\kappa  = \mathscr{P}\left(n_l,H;\kappa\right)\big\rvert_{\kappa_1}^{\kappa_{L/2}}\, .\label{eq:position_integral}
 \end{align}
The explicit representation of $\mathscr{P}$ in terms of the Elliptic integrals of the first and second kind is provided in Appendix \ref{app:elliptic_integrals}.
 The two transcendental equations \eqref{eq:two_point_contact_relations} can be numerically solved to obtain the values of $H$ and $\kappa_1$, which when substituted in \eqref{eq:angles_unknown} and \eqref{eq:arc_lengths_coordinates} deliver $\theta_1$ and $s_1$.
 Furthermore, the incoming tail of the configuration can be obtained by integrating \eqref{eq:configuration} with initial conditions \eqref{eq:incoming_initial_conditions}, while the symmetric left half of the lift-off region can be obtained by integrating \eqref{eq:configuration} from $s_1$ to $L/2$ with the following initial conditions,
 \begin{subequations}\label{eq:lift_off_initial_conditions}
 	\begin{align}
 		&\theta(s_1^+) = \theta_1\, ,\quad x(s_1^+) = -(1+r)\sin\theta_1\, ,\quad y(s_1^+) = (1+r)\cos\theta_1\, ,\\
 		&\kappa(s_1^+) = \kappa_1\, ,\quad n_1(s_1^+) = -\left(H - \frac{1}{2}B\kappa_1^2\right)\tan\theta_1\, ,\quad n_3(s_1^+) = H - \frac{1}{2}B\kappa_1^2\,.
 	\end{align}
 \end{subequations}
 The right half of the configuration can then be computed by symmetry.

\section{Frictionless contact in the long length limit}\label{sec:frictionless_contact_infinite_length}
In this section, we consider frictionless contact in a limit where the length $L$ of the rod is very large compared to the radius $R$ of the capstan.
\begin{figure}[t]
	\includegraphics[width=0.95\linewidth]{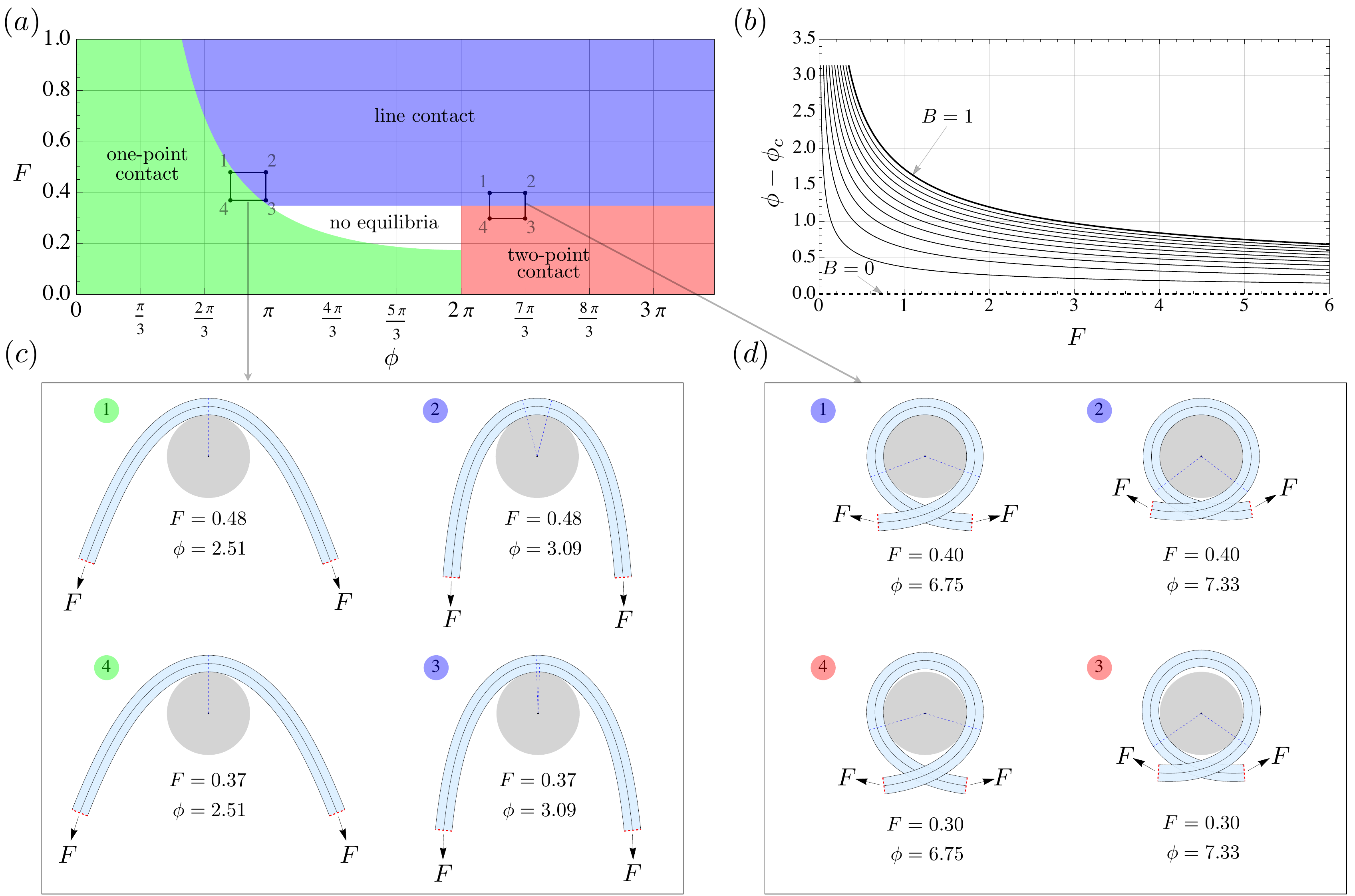}
	\caption{(a) Regions corresponding to one-point, line, and two-point contact in the $\{F,\phi\}$ space for a sufficiently long elastic rod with $B=1$. The curve at the intersection of the green and the blue region is given by equation \eqref{eq:point_contact_long_length} with $\kappa_1=\kappa_c$.
		Equilibrium configurations corresponding to the vertices of the two closed rectangular paths are shown in (c) and (d), where the red dashed caps at the two ends of the rod signify that the actual length of the rod is much longer than the displayed length.
		A video showing quasi-static transition of contact along the two loops is available in \cite{video}.
		(b) Difference between the wrap angle $\phi$ and the corresponding contact angle $\phi_c$ (for line contact), given by equation \eqref{eq:angle_difference}, is plotted against $F$ for different values of $B$. Each curve begins with a minimum value of $F$, given by $F_{min}=(1/2) B\kappa_c^2$, necessary to maintain line contact between the rod and the capstan. The curve at $B=0$ corresponds to a fully flexible filament for which $\phi=\phi_c$ for all values of $F$.  For a non-zero value of $B$ the difference $\phi-\phi_c$ approaches zero asymptotically as $F$ tends to infinity.}
	\label{fig:classification}
\end{figure}
We show that, in this limit, computations from the previous section undergo radical simplification, which allows for a clean classification of the three kinds of previously discussed equilibria.

Consider the shape equation \eqref{eq:curvature_equation} valid in the incoming and outgoing tails in a typical configuration.
Scaling it by the total length $L$ (normalized by $R$) of the rod, we obtain $(\tfrac{B}{L^2}\kappa')^2 + (\tfrac{B}{2L^2}\kappa^2 - H)^2 = |\bn|^2$.
If the length $L$ is large, the equation becomes a singular perturbation problem in the parameter $B/L^2$, indicative of the existence of boundary layers in its solutions \cite{hinch1991}.
Outside the boundary layer, or the ``outer region'', we substitute $B/L^2 = 0$ to obtain the following relation up to leading order,
\begin{align}
	H=|\bn|\, ,\label{eq:long_length_limit}
\end{align}
where the negative root of the equation has been discarded. 
Since this relation involves two quantities conserved throughout the length of the rod, it must hold true inside the boundary layer, or the ``inner region'', as well.
In this limit, most of the bending energy of the rod gets concentrated near the boundary of the contact region, while away from it, the rod is relatively straight.
Equation \eqref{eq:long_length_limit}, when combined with \eqref{eq:frictionless_Hamiltonian}, implies,
\begin{align}
	H = F_0=F_L=F\, ,\label{eq:frictionless_long_length_end_loads}
\end{align}
meaning that the applied end loads must be equal in order to maintain equilibrium.
In other words, asymmetric equilibria of the kind shown in Fig. \ref{fig:assymetric_equilibria} is not possible in the long length limit.

Another consequence of \eqref{eq:long_length_limit} is that the length constraints, given by \eqref{eq:point_contact_length_relation}, \eqref{eq:line_contact_relations}, and \eqref{eq:two_point_contact_length_relation}, become identically satisfied, which greatly simplifies the computations of the free boundaries in all the three kinds of equilibrium, as is shown next.

To compute one-point contact equilibria, we use equation \eqref{eq:point_contact_angle_relation} with \eqref{eq:frictionless_long_length_end_loads} and obtain after some rearrangement the following expression for the rod's centerline curvature at the contact point
\begin{align}
	\kappa_1 = 2\sqrt{\frac{F}{B}} \sin\frac{\phi}{4}\, .\label{eq:point_contact_long_length}
\end{align}
The region in the $\{F,\phi\}$ space with $\kappa_1\le\kappa_c$ corresponds to one-point contact equilibria, and is shown in green in Fig. \ref{fig:classification}.
The location of the point of contact can be deduced from symmetry as $\theta_1=0$, which can also be confirmed by substituting \eqref{eq:point_contact_long_length} in \eqref{eq:angles_unknown}$_1$.
The entire configuration can then be computed following the same procedure as described towards the end of Sect. \ref{sec:one_point_finite_length}.

When $\kappa_1$ delivered by equation \eqref{eq:point_contact_long_length} hits the critical curvature $\kappa_c$ for prescribed values of $F$ and $\phi$, one-point contact transitions to line contact.
In that case, $\theta_1$ and $\theta_2$ can be directly computed from \eqref{eq:angles_unknown} using equation \eqref{eq:frictionless_long_length_end_loads}, along with the fact that $\kappa=\kappa_c$ in the contact region. 
Consequently, the contact angle $\phi_c=\theta_1-\theta_2$ can be explicitly written as
\begin{align}
	 \phi_c = \phi - 2\arccos\left(1-\frac{B}{2F} \kappa_c^2\right),\quad\text{with}\quad F>\frac{1}{2}B\kappa_c^2\, ,\label{eq:angle_difference}
\end{align}
where the inequality ensures that the contact pressure from equation \eqref{eq:line_contact_shaer_tension_pressure}, i.e., $p_1=F\kappa_c - (1/2)B\kappa_c^3$, remains non-negative.

The region corresponding to line contact in the $\{F,\phi\}$ space, with $p_1>0$, is shown in blue in Fig. \ref{fig:classification}a.
No contact equilibria exists for the region shown in white\footnote{Note that for a given value of $B$, one can always find a finite length of the rod for which a one-point equilibrium exists at any chosen point in the white region in Fig. \ref{fig:classification}a. }, as the contact pressure between the capstan and the rod in that region becomes negative.
The difference $\phi-\phi_c$ from \eqref{eq:angle_difference} is plotted against $F$ for different values of $B$ in Fig. \ref{fig:classification}b.
The shear force, tension, and pressure in the contact region can be computed using \eqref{eq:line_contact_shaer_tension_pressure},
and the incoming and outgoing tail can be constructed using the same procedure as described in the end of Sect.  \ref{sec:line_contact_finite_length}.

For $\phi>2\pi$, and $F<\tfrac{1}{2}B \kappa_c^2$, the pressure required to maintain line contact between the rod and the capstan becomes negative, leading to line contact degenerating into two-point contact.
In Fig. \ref{fig:classification}, the corresponding region is shown in red.
To compute such configurations, we obtain $\theta_1$ in terms of the unknown curvature $\kappa_1$ at the point of contact using equation \eqref{eq:angles_unknown}$_1$ and \eqref{eq:frictionless_long_length_end_loads}.
Thereafter, $\kappa_1$ can be obtained by solving \eqref{eq:two_point_contact_position_relation} numerically.
Finally, the full configuration can be constructed as described towards the end of Sect. \ref{sec:two_point_finite_length}.

 \section{Frictional contact and limiting equilibrium} \label{sec:frictional_contact}
We now address the last facet of the generalized capstan problem, i.e., a generalization of the classic capstan equation \eqref{eq:classic_capstan}, governing limiting equilibria of an elastic rod in frictional contact with a circular capstan.
Before we proceed further, a few comments on the problem are in order.

We showed in Sect. \ref{sec:force_amplification_frictionless} that force amplification across a finite length of an elastic rod wrapped around a circular capstan can be achieved in the absence of frictional interactions between them.
Since friction appears unnecessary to support unequal loads at the two ends of the rod, a precise definition of limiting equilibrium seems unclear.
However, in the long length limit discussed in Sect. \ref{sec:frictionless_contact_infinite_length}, relation \eqref{eq:frictionless_long_length_end_loads} forbids frictionless equilibria with unequal end loads, which insinuates that friction is indeed necessary for force amplification in this limit.
Therefore, we will pursue the generalization of the classical capstan equation \eqref{eq:classic_capstan} in the long length limit, where we define limiting equilibrium under frictional contact as the configuration which maximizes the output load $F_L$, for a given input load $F_0$ and wrap angle $\phi$.
We will also consider only line contact between the rod and the capstan, as corollary \eqref{eq:Hamiltonian_rewritten} of our assumption \eqref{eq:Hamiltonian_jump_condition} forbids any point reaction force $P_3$ in the tangential direction at isolated points of contact.

\begin{figure}[t]
	\includegraphics[width=0.95\linewidth]{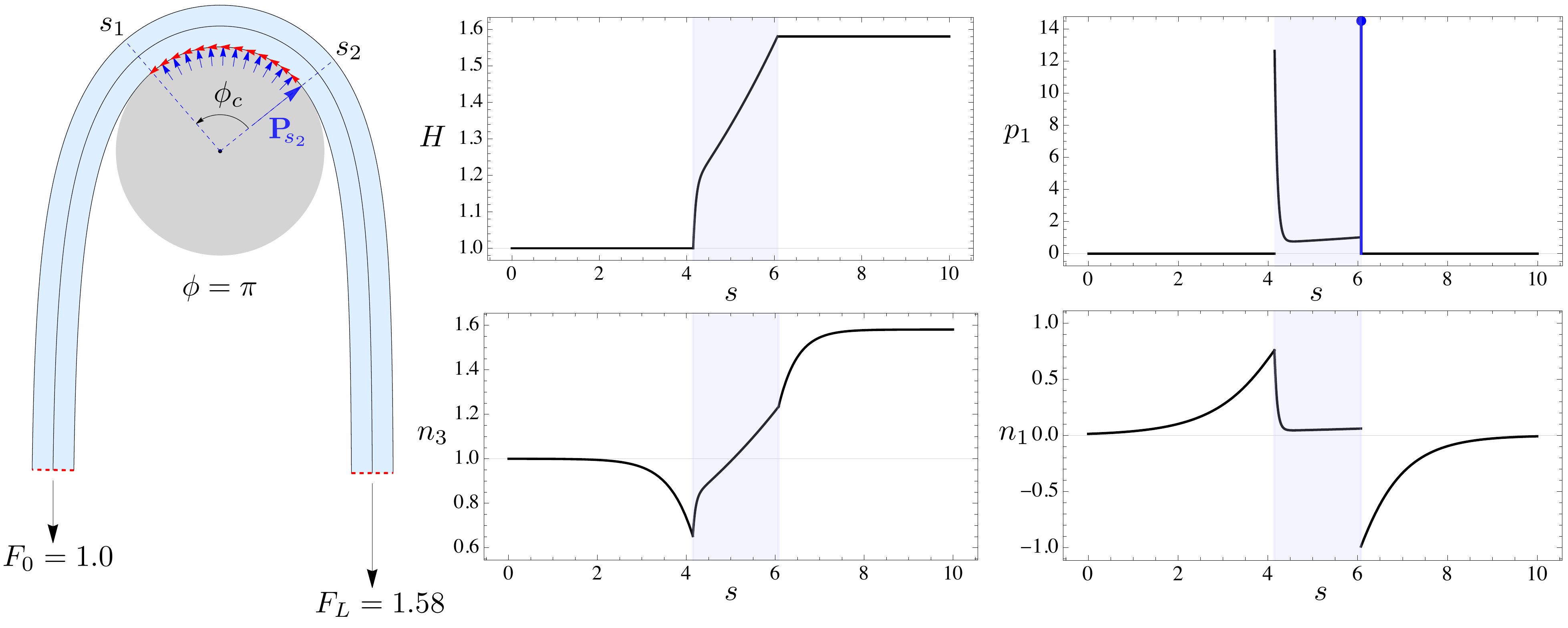}
	\caption{On the left is a configuration corresponding to limiting equilibrium of a sufficiently long elastic rod in frictional contact with a rigid capstan with $\phi_c=1.61$. The coefficient of friction $\mu=0.3$. The rod experiences a normal point force of magnitude $|\bP_{\!\! s_2}|=1.05$ at $s_2$, while no such point reaction force exists at $s_1$ for reasons explained in the text surrounding equation \eqref{eq:shear_jump_frictional_capstan}. The plots of the Hamiltonian $H$, normal pressure $p_1$, tension $n_3$, and shear force $n_1$ are also shown. The shaded region on the plots corresponds to the contact region. The normal pressure $p_1$ suffers a Dirac-delta function singularity at the outgoing end of the contact region, causing a jump in the shear force $n_1$ at the same point.}
	\label{fig:frictional_capstan}
\end{figure}

Consider a sufficiently long elastic rod wrapped around a circular capstan as shown in Fig. \ref{fig:frictional_capstan}.
Using equations \eqref{eq:long_length_limit} and \eqref{eq:Hamiltonian_conservation} in the incoming and outgoing tails, along with \eqref{eq:Hamiltonian_jump_condition} across the contact boundaries $s_1$ and $s_2$, we can establish the following relations,
\begin{align}
	H(s_1^-) = H(s_1^+) = F_0\quad\text{and}\quad H(s_2^-) =  H(s_2^+)=F_L\, .\label{eq:Hamiltonian_and_force_relations}
\end{align}
With the second relation above, the requirement of maximizing the output load $F_L$ can be transferred to maximizing the value of $H(s_2)$.
The evolution of the Hamiltonian function $H$ of the rod in the contact region is governed by the following equations obtained by substituting $\kappa = \kappa_c$ in \eqref{eq:Hamiltonian_and_friction_ND},
\begin{subequations}\label{eq:frictional_contact_general}
\begin{align}
	H' &= p_3(1-r\kappa_c)\, ,\label{eq:Hamiltonian_frictional_general}\\
	rp_3' &= H\kappa_c - p_1 - \frac{1}{2}B\kappa_c^3 \, .\label{eq:friction_general}
\end{align}
\end{subequations}

The two ODEs above contain three unknown functions of $s$, namely $H$, $p_3$, and $p_1$.
To obtain a closed system of equations, we assume frictional interaction between the rod and the capstan to be governed by Coulomb's inequality of static friction, which relates the frictional force density $p_3$ and the normal pressure $p_1$ as,
\begin{align}
	p_3\le\mu p_1\, ,\label{eq:friction_inequality}
\end{align}
where $\mu$ is the coefficient of static friction between the rod and the capstan.

To maximize the value of $H(s_2)$, and consequently of the output load $F_L$, we seek those solutions of \eqref{eq:frictional_contact_general} for which $H$ grows at the fastest possible rate over the yet unknown contact region.
This is ensured when inequality \eqref{eq:friction_inequality} saturates, and $p_3$ attains its maximum value given by $p_3=\mu p_1$.
\begin{figure}[t]
	\includegraphics[width=0.95\linewidth]{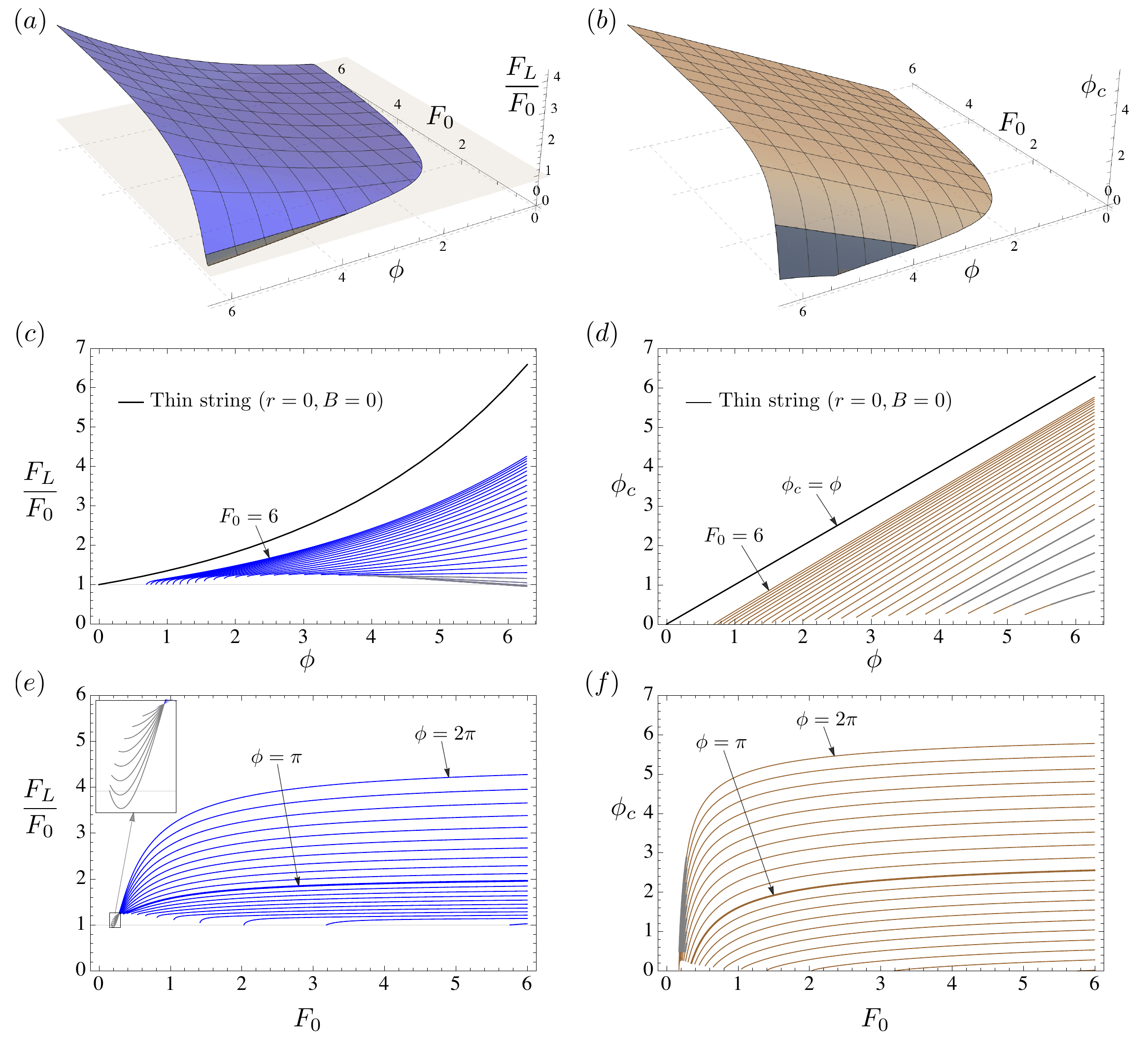}
	\caption{(a) A 3D plot of maximum force ratios $F_L/F_0$ as a function of $F_0$ and $\phi$, for a sufficiently long rod with $B=1$, and coefficient of friction $\mu=0.3$. The gray region of the surface corresponds to configurations where negative pressures between the rod and the capstan develop, and are therefore unphysical. The two perspectives of the surface in the $\{F_L/F_0,\phi\}$ and $\{F_L/F_0,F_0\}$ plane are shown in (c) and (e) respectively.  (b) A 3D plot of the contact angle $\phi_c$ as a function of $F_0$ and $\phi$. As in (a), the gray region corresponds to configurations where negative pressures develop. The two perspectives of the surface in the $\{\phi_c,\phi\}$ and $\{\phi_c,F_0\}$ plane are shown in (d) and (f) respectively.}
	\label{fig:surfaces}
\end{figure}
Using this equality, we eliminate $p_3$ in favor of $p_1$ in \eqref{eq:frictional_contact_general} to obtain the following set of equations governing limiting equilibria of the rod,
\begin{subequations}\label{eq:frictional_contact_final}
	\begin{align}
		H' &= \mu p_1(1-r\kappa_c)\, ,\label{eq:Hamiltonian_frictional_final}\\
		r\mu p_1' &= H\kappa_c - p_1 - \frac{1}{2}B\kappa_c^3 \, .\label{eq:friction_final}
	\end{align}
\end{subequations}
The above set of ODEs, which are linear in $H$ and $p_1$, can be integrated analytically to obtain,
\begin{subequations}\label{eq:frictional_contact_solution}
\begin{align}
	H(s) &= c_1 \left(\frac{A+1}{2\kappa_c}\right)e^{\frac{A-1}{2r\mu}(s-s_1)} - c_2\left( \frac{A-1}{2\kappa_c}\right)e^{-\frac{A+1}{2r\mu}(s-s_1)} + \frac{1}{2}B\kappa_c^2\, ,\label{eq:Hamiltonian_solution}\\
	p_1(s)&=c_1 e^{\frac{A-1}{2r\mu}(s-s_1)} + c_2 e^{-\frac{A+1}{2r\mu}(s-s_1)}\, ,\label{eq:pressure_solution}
\end{align}
\end{subequations}
where $s\in[s_1,s_2]$, $A = \sqrt{1+4r\kappa_c\mu^2(1-r\kappa_c)}$, and $c_1$ and $c_2$ are constants given by,
\begin{align}
	c_1 = \frac{p_1(s_1^+)(A-1) + \kappa_c\left[2H(s_1^+) - B\kappa_c^2\right]}{2A}\,, \quad
	c_2 = \frac{p_1(s_1^+)(A+1) - \kappa_c\left[2H(s_1^+)- B\kappa_c^2\right]}{2A}\, .\label{eq:constants_of_integration}
\end{align}
These constants must be determined by the initial conditions $p(s_1^+)$ and $H(s_1^+)$ accompanying the system \eqref{eq:frictional_contact_final}.

To compute $p(s_1^+)$, we invoke the jump condition \eqref{eq:jump_conditions_director_components}$_2$ on the shear force at $s_1$, and note that $p_1(s_1^-) = 0$, $p_3(s_1^+)=\mu p_1(s_1^+)$, and $\kappa'(s_1^+) = 0$.
The said jump condition can then be rearranged as,
 \begin{align}
	r\mu p_1(s_1^+)  = B\kappa'(s_1^-) - P_1(s_1)\, .\label{eq:shear_jump_frictional_capstan}
\end{align}
Both $p_1$ and $P_1$ in the equation above are positive numbers -- as the capstan can only push on the rod and not pull -- and as a consequence, $\kappa'(s_1^-)$ must also be positive.
To maximize the growth rate of $H$ at $s_1^+$, the highest possible value of $p_1(s_1^+)$ must be deduced from \eqref{eq:shear_jump_frictional_capstan}, which is ensured when $P_1(s_1) = 0.$\footnote{This argument has precedence in \cite{grandgeorge2021}, where limiting equilibria of a fully flexible filament of finite thickness in contact with a curved capstan was considered.}
This implies that at limiting equilibria no point reaction force can occur at $s_1$.\footnote{The same conclusion in the context of an elastic rod in frictional contact with a rotating cylinder was reached in \cite{grandgeorge2022}.  However, at variance with the present approach, the authors there employ a ``geometrical argument'', where they invoke the circular geometry of the capstan to arrive at this conclusion.}
Consequently, we have $p_1(s_1^+)=B\kappa'(s_1^-)/r\mu$ from \eqref{eq:shear_jump_frictional_capstan},
where the term $B\kappa'_1(s_1^-) = (\kappa_c/2)\sqrt{B[4H(s_1^-) - B\kappa^2(s_1^-)]}$ can be obtained from \eqref{eq:curvature_equation} applied at $s=s_1^-$.
Finally, noting the equalities at $s_1$ from \eqref{eq:Hamiltonian_and_force_relations}, along with $\kappa(s_1^-) = \kappa(s_1^+)=\kappa_c$ from \eqref{eq:tension_curvature_jump}$_2$, we can write the initial conditions for \eqref{eq:Hamiltonian_frictional_final} at $s=s_1^+$ as,
  \begin{align}
	H(s_1^+) = F_0\,,\qquad 	p_1(s_1^+) = \frac{\kappa_{c}}{2r\mu}\sqrt{B (4F_0 - B\kappa_c^2)}\, .\label{eq:Hamiltonian_pressure_initial_condition}
\end{align} 
The constants $c_1$ and $c_2$ in \eqref{eq:frictional_contact_solution} can now be determined by substituting \eqref{eq:Hamiltonian_pressure_initial_condition} in \eqref{eq:constants_of_integration}, thereby completing the description of $H(s)$ and $p_1(s)$ in the contact region.
The normal reaction force at the outgoing boundary $s_2$ can be computed using \eqref{eq:jump_conditions_director_components}$_2$, along with $p(s_2^+) = 0$ and $\kappa'(s_2^-) = 0$, to be $P_1(s_2) = r\mu p_1(s_2^-) - B\kappa'(s_2^+)$.
The contact region between the rod and the capstan still remains an unknown, which we compute next.

The extent of contact region is measured by the length $s_2-s_1$.
Since the length of the rod is infinitely large, the precise value of $s_1$ is irrelevant, and can be assigned an arbitrary value for simplicity leaving $s_2$ to be the only unknown to be found.
We note that $s_2-s_1 = (1+r)(\theta_1-\theta_2)$, and substitute in it $\theta_1$ and $\theta_2$ from \eqref{eq:angles_unknown}.
In the resulting expression we use $H_0=F_0$ and $H_L=F_L=H(s_2)$ due to \eqref{eq:Hamiltonian_and_force_relations} and \eqref{eq:Hamiltonian_conservation}, and $\kappa_1=\kappa_2=\kappa_c$ due to \eqref{eq:tension_curvature_jump}$_2$ across $s_1$ and $s_2$, to obtain the following equation after some rearrangement,
\begin{align}
	\arccos\left(1-\frac{B\kappa_c^2}{2F_0}\right)+\frac{s_2-s_1}{1+r}+\arccos\left(1-\frac{B\kappa_c^2}{2H(s_2)}\right) = \phi\, .\label{eq:angle_contact}
\end{align}
The first and the third term on the left measure the angular displacements of the tangents between the two terminal ends and their corresponding contact boundaries, while the middle term measures the angular displacement of the tangent in the contact region.
The above equation essentially constrains the angular displacements of the tangents in the tails and the contact region to add up to the prescribed wrap angle $\phi$.
The expression for $H(s_2)$ can be substituted in \eqref{eq:angle_contact} from \eqref{eq:Hamiltonian_solution}, and the entire equation can be numerically solved to obtain $s_2$.
The output load corresponding to limiting equilibrium is then given by $F_L=H(s_2)$, and the contact angle can be computed as $\phi_c = (s_2-s_1)/(1+r)$.

The surface generated by the maximum force ratios $F_L/F_0$ for an elastic rod in the $\{F_L/F_0,F_0,\phi\}$ space is plotted in Fig. \ref{fig:surfaces}a for $B=1$ and $\mu=0.3$.
The corresponding surface for the contact angles $\phi_c$ is shown in Fig. \ref{fig:surfaces}b.
The gray region marked on both the surfaces represents configurations with small initial force and high wrap angle where the contact pressure $p_1$ goes negative, and must therefore be discarded.
Note that unlike in the classic capstan problem, the force ratio $F_L/F_0$ in the presence of bending elasticity is also a function of the initial force $F_0$, as can be seen in the projection of the surface in Fig. \ref{fig:surfaces}e.
A comparison between the classic and the generalized capstan problem in Fig. \ref{fig:surfaces}c shows that force ratios for the latter are always lower than the former. 
The solutions to the generalized problem asymptotically merge with the classic solution as the input force $F_0$ tends to infinity.
A typical configuration of a sufficiently long elastic rod in frictional contact with a circular capstan is shown in Fig. \ref{fig:frictional_capstan}, along with the plots of the Hamiltonian function $H$, contact pressure $p_1$, tension $n_3$, and shear force $n_1$, against the arc-length coordinate.

\section{Conclusion}\label{sec:conclusion}
We have presented a generalization of the classic capstan problem to include the effects of finite thickness and bending elasticity of the filament.
The problem was referred to as the \emph{generalized capstan problem}, where we modeled the filament as an elastic rod of finite thickness, and computed configurations with both frictionless and frictional contact between the rod and the capstan.
Unlike much of the prior work on the subject, we treated the boundaries of the contact region as \emph{free boundaries} \cite{burridgekeller1978,oreilly2017}, and computed their location with the aid of a jump condition, systematically derived using the principle of virtual work.

For frictionless contact, we computed three kinds of rod equilibria, where the rod touches the capstan at one point, along a continuous arc, and at two points connected by an intermediate lift-off (or contact-free) region.
Our analysis revealed that, in general, a finite length of elastic rod wrapped around a circular capstan does not require friction to sustain unequal loads at its two ends.
This is in stark contrast to the classic case governed by equation \eqref{eq:classic_capstan}, where friction is necessarily required for force amplification across the length of the filament.
We showed this property of elastic rods to be a consequence of the conservation of the \emph{Hamiltonian function} throughout its length.
We then considered a limit where the length of the rod is much larger than the radius of the capstan.
In this limit, we showed that frictionless contact equilibria between the capstan and the rod with unequal end loads is no longer possible.
This long length limit lead to significant simplifications in the computation of frictionless equilibria, and allowed for a clean classification of the three aforementioned kinds of frictionless contact equilibria in a two parameter space of the applied end load $F$ and wrap angle $\phi$.

Finally, we incorporated frictional interaction between the elastic rod and the capstan under the assumption of the long length limit, and obtained a generalization of the classic capstan equation \eqref{eq:classic_capstan} to include bending elasticity and finite thickness of the rod.
We showed that the maximum force ratios predicted by our model depend not only on the wrap angle $\phi$, but also on the prescribed input load $F_0$, which is in contrast to the classic capstan problem where the force ratios are independent of the latter.
We found that for any given input load $F_0$ and wrap angle $\phi$, the maximum force ratios predicted by our model remained lower than the predictions made by the classic capstan equation \eqref{eq:classic_capstan}.
Since our theory inputs the wrap angle $\phi$ as opposed to the contact angle $\phi_c$ -- which is much more difficult to enforce experimentally than the former -- we hope that its validity could be put to test against simple experiments.

\section{Acknowledgments}
The author is grateful to John H. Maddocks for several helpful discussions on every aspect of this work. He also acknowledges useful discussions with Paul Grandgeorge. This work was partially supported by Swiss National Science Foundation Grant 200020-182184 to John H. Maddocks.

\section{Conflict of interest statement}
The author has no conflict of interest to declare.

\appendix
\section{Derivation of the free boundary condition}\label{app:contact_condition_derivation}
Here, following the procedure laid out in \cite{hill1951}, we present the details of the derivation resulting in equation \eqref{eq:virtual_work_localised}.
Consider an infinitesimal transformation in the arc-length coordinate $s\rightarrow s^*$, with $s^*$ given by equation \eqref{eq:arc_length_variation}, such that the corresponding total variation in the position vector and the directors is zero, i.e. $\delta\bx(s) \equiv \bx^*(s^*)-\bx(s) = 0$, and $\delta\bd_i(s)\equiv \bd^*_i(s^*)-\bd_i(s)=0$.
We then define a variation $\tilde{\delta}$ in the position vector and the directors at a material label $s$ such that,
\begin{align}
	\bx^*(s) := \bx(s)  + \tilde{\delta}\bx(s)\, ,\quad\bd_i^*(s) := \bd_i(s) + \tilde{\delta}\bd_i(s)\, .\label{eq:variations_at_labels}
\end{align}
Using the conditions $\bx^*(s^*)=\bx(s)$ and $\bd^*_i(s^*)=\bd_i(s)$, the variations $\tdelta\bx(s)$ and $\tdelta\bd_i(s)$ can be represented as,
\begin{align}
	\tdelta\bx(s)  =- \bx'(s)\delta s\, ,\quad \tdelta\bd_i(s) =  - \bd_i'(s)\delta s\, .\label{eq:variations}
\end{align}
To compute the corresponding variation in the Darboux vector, we observe that,
\begin{align}
(\bd_i^*)' = \bu^*\times\bd_i^* = \bd_i' + \bu\times\tdelta\bd_i + \tdelta\bu\times\bd_i + \text{higher order terms.}\, \label{eq:Darboux_variation_computation}
\end{align}
The orthonormality of the director frame requires 
\begin{align}
\tdelta\bd_j = \tilde\delta\bz\times\bd_j\, ,\label{eq:variation_orthonormality}
\end{align}
 where $\tdelta\bz$
 is some infinitesimal function of the independent coordinates.
 Using \eqref{eq:variations}$_2$ and \eqref{eq:centreline_and_director_evolution}$_2$ in \eqref{eq:variation_orthonormality}, $\tilde\delta\bz$ can be represented as,
 \begin{align}
 	\tilde\delta\bz = -\bu\,\delta s\, .\label{eq:delta_z}
 \end{align}
Substituting \eqref{eq:variation_orthonormality} into \eqref{eq:Darboux_variation_computation}, and using the fact that $\bd_i$ is a basis of $\mathbb{R}^3$, we obtain after some manipulations the relation $\tdelta\bu = \tdelta\bz ' - \bu\times\tdelta\bz$, from which we deduce
\begin{align}
\tdelta u_i = \tdelta \bz'\cdot\bd_i \label{eq:Darboux_variation_components}
\end{align}
where $\tdelta u_i\equiv\tdelta(\bu\cdot\bd_i)$.

Following standard procedure \cite{hill1951}, and using \eqref{eq:variations} and \eqref{eq:Darboux_variation_components}, the variation due to $s\rightarrow s^*$ in the energy functional \eqref{eq:energy} can be computed as,

\begin{align}
	\tdelta\mathcal{E} = \int_{s_1}^{s_2}\left[ \left(W(\bu,\bx) - \bn\cdot\bx' - \bmo\cdot\bu\right)\delta s \right]'ds  + \int_{s_1}^{s_2}\left[-\left(\frac{\partial W}{\partial\bu}\right)' - \bd_3\times\bn\right]\cdot\tdelta\bz \,ds+ \int_{s_1}^{s_2}\left[-\bn'\right]\cdot\tdelta\bx \,ds\, .\label{eq:work_internal}
\end{align}
Finally, we substitute the above, along with $\tdelta W_{ext}$ from \eqref{eq:work}, in \eqref{eq:virtual_work_statement}, and then split various integrals into two domains $[s_1,s_0^-]$ and $[s_0^+,s_2]$.
Thereafter, we use \eqref{eq:force_and_moment_local_balance} and \eqref{eq:linear constitutive} to get rid of the bulk terms in the two domains.
The remaining terms are localised around $s=s_0$ by taking the limit $s_1\rightarrow s_0^-$ and $s_2\rightarrow s_0^+$ to arrive at \eqref{eq:virtual_work_localised}.

\section{Elliptic integrals}\label{app:elliptic_integrals}
Consider the length integral stated in \eqref{eq:length_function}.
A change of variable from $\kappa$ to $t$, such that $t=\sqrt{B/2(H-|\bn|)}\kappa$, transforms the integral into $i\sqrt{2B/(H+|\bn|)}\int\left[(1-t^2)(1-k t^2)\right]^{-1}dt$, which can subsequently be transformed to a canonical form using $t=\sin\theta$.
The resulting expression is,
\begin{align}
	\mathscr{L}(|\bn|,H; \kappa) := -i\sqrt{\frac{2B}{H+|\bn|}}\, F\left(\varphi \,|\, k\right)\, ,
\end{align}
where $F(\varphi\,|\, k)$ is the Elliptic Integral of the first kind \cite{abramowitz1972,byrdfriedman1971} given by, 
\begin{align}
	F\left(\varphi \,|\, k\right) = \int_0^\varphi\frac{d\theta}{\sqrt{1-k\sin^2\theta}}\, ,\qquad\varphi = \arcsin\left(\sqrt{\frac{B}{2(H-|\bn|)}}\kappa\right)\, ,\qquad k = \frac{H-|\bn|}{H+|\bn|}.\label{eq:elliptic_first_kind}
\end{align}
Using the same change of variables as for the length integral from $\kappa$ to $t$ and then from $t$ to $\theta$, the integral in \eqref{eq:position_integral} can be written as,
\begin{align}
	\mathscr{P}(|\bn|,H: \kappa) := -i\sqrt{\frac{2B}{H+|\bn|}}\left[(H+|\bn|)E\left(\varphi \,|\, k\right) - |\bn| F\left(\varphi \,|\, k\right)\right]\, ,
\end{align}
where $F(\varphi\,|\,k)$ is given by \eqref{eq:elliptic_first_kind}, while $E(\varphi\,|\,k)$ is the Elliptic Integral of the second kind \cite{abramowitz1972,byrdfriedman1971} given by,
\begin{align}
E\left(\varphi \,|\, k\right) = \int_0^\varphi\sqrt{1-k\sin^2\theta}\,d\theta\, ,
\end{align}
with the same expressions for $\varphi$ and $k$ as stated in \eqref{eq:elliptic_first_kind}.

Note that although $\mathscr{L}$ and $\mathscr{P}$ are in general complex valued functions, their differences between any two values of $\kappa$ (for fixed $|\bn|$ and $H$), i.e.  $\mathscr{L}(|\bn|,H;\kappa)|_{\kappa_A}^{\kappa_B}$ and $\mathscr{P}(|\bn|,H;\kappa)|_{\kappa_A}^{\kappa_B}$, are always real valued.
This is due to the fact that the real components $F(\varphi\,|\, k)$ and $E(\varphi\,|\,k)$ (or the imaginary components of $\mathscr{L}$ and $\mathscr{P}$) are functions of $k$ alone \cite{byrdfriedman1971}, and get canceled out when subtracted with different values of $\varphi$ (or $\kappa$).


\end{document}